\documentclass[aps,prd,amsmath,floats,floatfix,twocolumn,superscriptaddress,nofootinbib]{revtex4-1}
\usepackage{graphicx,amssymb,amsmath,amsbsy,mathrsfs}
\usepackage{bm}

\usepackage[usenames]{color}
\usepackage{ulem}
\newcommand{\beq}{\begin{equation}}
\newcommand{\eeq}{\end{equation}}
\newcommand{\beqa}{\begin{eqnarray}}
\newcommand{\eeqa}{\end{eqnarray}}

\newcommand{\SMetric}{g}     
\newcommand{\SRicci}{R}      
\newcommand{\SRicciS}{R}     

\newcommand{\ExCurv}{K}      
\newcommand{\TrExCurv}{K}    

\newcommand{\CF}{\psi}              

\newcommand{\CMetric}{{\tilde{g}}}    

\newcommand{\Spin}{J} 

\newcommand{\HorizonMass}{M}

\newcommand{\Caltech}{\affiliation{Theoretical Astrophysics 350-17,
    California Institute of Technology, Pasadena, California 91125}}

\newcommand{\CITA}{\affiliation{Canadian Institute for Theoretical Astrophysics,
60 St. George Street, University of Toronto, Toronto, ON M5S 3H8, Canada}}

\begin{document}
\vspace{-2.5cm} 

\title{Horizon dynamics of distorted rotating black holes}

\author{Tony Chu} \Caltech
\author{Harald P. Pfeiffer} \CITA
\author{Michael I. Cohen} \Caltech

\date{\today}

\begin{abstract}
    We present numerical simulations of a rotating 
    black hole distorted by a pulse of ingoing 
    gravitational radiation.  For strong pulses, 
    we find up to five concentric marginally
    outer trapped surfaces.  These trapped surfaces appear and disappear in
    pairs, so that the total number of such surfaces at any given
    time is odd.  The world tubes traced out by the marginally outer
    trapped surfaces are found to be spacelike during the highly dynamical 
    regime, approaching a null hypersurface at early and late times. 
    We analyze the structure of these marginally trapped tubes in the context 
    of the dynamical horizon formalism, computing the expansion of outgoing and
    incoming null geodesics, as well as evaluating the dynamical
    horizon flux law and the angular momentum flux law.  Finally, 
    we compute the event horizon.  The event horizon is well-behaved and
    approaches the apparent horizon before and after the highly
    dynamical regime.  No new generators enter the event horizon
    during the simulation.
\end{abstract}

\maketitle

\section{Introduction}

In the efforts by the numerical relativity community leading up to the 
successful simulation of the inspiral and merger of two black holes, 
analyses of single black holes distorted by gravitational radiation have 
offered a convenient and simpler setting to understand the nonlinear dynamics 
during the late stages of binary black hole coalescence. For this purpose, 
initial data for a Schwarzschild black hole plus a Brill wave was presented 
in~\cite{ancsa94}, which was both time symmetric and axisymmetric. 
In highly distorted cases, the apparent horizon could develop very long, 
spindlelike geometries. If the event horizon can show similar behavior, 
this would raise intriguing questions related to the hoop 
conjecture~\cite{Thorne72}. The work of~\cite{ancsa94}
was extended to distorted rotating black holes in~\cite{AnninosEtAl:1994}, 
where the apparent horizon served as a useful tool to examine the quasinormal 
oscillations of the black hole geometry as it relaxed in an evolution. 
Further studies have extracted the gravitational waves emitted by the 
black hole~\cite{BrandtSeidel:1995}, and compared the apparent and event 
horizons~\cite{Libson95a}.

We continue this line of investigation here, 
while incorporating various modern notions of quasilocal horizons that 
have emerged in recent years. Our emphasis is on horizon properties 
during the highly dynamical regime, and no symmetries  
are present in our initial data and evolutions.
The utility of quasilocal horizons can be immediately appreciated when 
one wants to perform a numerical evolution of a black hole spacetime. 
One must be able to determine the surface of the black hole at each time, 
in order to track the black hole's motion and compute its properties, 
such as its mass and angular momentum. However, the event horizon, which is 
the traditional notion of a black hole surface, can only be found after 
the entire future history of the spacetime is known.

Quasilocal horizons can be computed locally in time, and so are used instead 
to locate a black hole during the evolution.
Of particular interest is a marginally outer trapped surface (MOTS), which 
is a spatial surface on which the expansion of its outgoing null normal 
vanishes~\cite{hawkingellis}. 
The use of MOTSs is motivated by several results. When certain positive 
energy conditions are satisfied, an MOTS is either inside of or coincides 
with an event horizon~\cite{hawkingellis,Wald}. The presence of an MOTS 
also implies the existence of a spacetime singularity~\cite{Penrose1965b}.
In an evolution, the MOTSs located at successive times foliate a world tube, 
called a marginally trapped tube (MTT). MTTs have been studied in the 
context of trapping horizons~\cite{Hayward1994,Hayward2004}, 
isolated horizons~\cite{Ashtekar2000,Ashtekar2000_Generic,Ashtekar2001}, 
and dynamical horizons~\cite{Ashtekar2002,Ashtekar2003,Ashtekar2005}.

Both the event horizon and an MTT react to infalling matter and radiation, 
although their behaviors can be quite different in highly dynamical 
situations. Being a null surface, the evolution of the event horizon is  
governed by the null Raychaudhuri equation~\cite{PoissonToolkit}, 
so that even though 
its area never decreases, in the presence of infalling 
matter and radiation the rate of growth of its area decreases and can 
even become very close to zero~\cite{Booth2005}. Since an MTT 
is determined by quasilocal properties of the spacetime, its 
reaction to infalling matter and radiation is often much more intuitive. 
A MTT is usually spacelike (e.g. a dynamical horizon) in such 
situations, although further scrutiny has revealed that MTTs can exhibit  
various intriguing properties of their own. For example, an MTT may become 
timelike and decrease in area~\cite{Booth2006}, or even have sections that 
are partially spacelike and partially timelike~\cite{Schnetter2006}. 
In a numerical simulation, such behavior is often indicated by the 
appearance of a pair of new MTTs at a given time, accompanied by a 
discontinuous jump in the world tube of the apparent horizon, or 
outermost MOTS.  

In this paper, we investigate the behavior of MTTs and the event horizon 
in the context of a rotating black hole distorted by an ingoing pulse of 
gravitational waves. First, we construct a series of initial data sets in 
which the amplitude of the gravitational waves varies from small to large, 
which are then evolved. We focus on the evolution with the largest 
distortion of the black hole, in which the mass of the final black hole 
is more than double its initial value. During the evolution, the world tube 
of the apparent horizon jumps discontinuously when the gravitational waves 
hit the black hole, and as many as five MTTs are found at the same time. 
Some of these MTTs 
decrease in area with time, although we find that all the MTTs during the 
dynamical stages of our evolution are spacelike and dynamical horizons. 
Moreover, all these MTTs join together as a single dynamical horizon. 
Their properties are further analyzed using the dynamical horizon flux 
law~\cite{Ashtekar2003}, which allows one to interpret the growth of the 
black hole in terms of separate contributions. 
We also evaluate the angular momentum flux law based on the generalized 
Damour-Navier-Stokes equation~\cite{Gourgoulhon2005b}.
Finally, we locate the event horizon and contrast its behavior with that of 
the MTTs.

The organization of this paper is as follows. Section II details the 
construction of the initial data sets and Sec. III describes the 
evolutions. Section IV introduces some definitions about MOTSs, and the 
methods used to locate them. 
Section V discusses the MTTs foliated by the MOTSs, the determination of their 
signatures, and the fluxes of energy and angular momentum across them. 
The emphasis is on the case with the largest 
distortion of the initial black hole, as is the remainder of the paper. 
Section VI explains how we find the event horizon, and contrasts its 
properties with the MTTs. Section VII presents some concluding 
remarks. Finally, the appendix offers some insight on our results
in light of the Vaidya spacetime.

\section{Initial Data}
\label{sec:ID}
Initial data sets are constructed following 
the method of~\cite{Pfeiffer2004}, which is based on the extended 
conformal thin sandwich formalism. First, the 3+1 decomposition of the
spacetime metric is given by~\cite{ADM,york79}
\begin{align}
\label{eq:3Plus1Metric}
^{(4)}ds^2 &= g_{\mu\nu}dx^{\mu}dx^{\nu},\\
      &= -N^2dt^2+g_{ij}\left(dx^i+\beta^idt\right)\left(dx^j+\beta^jdt\right),
\end{align}
where $g_{ij}$ is the spatial metric of a $t=\text{constant}$ hypersurface 
$\Sigma_t$, $N$ is the lapse function, and $\beta^i$ is the shift vector. 
(Here and throughout this paper, Greek indices are spacetime indices running 
from 0 to 3, while Latin indices are spatial indices running from 1 to 3.)
Einstein's equations (here with vanishing stress-energy tensor $T_{\mu\nu}=0$) 
then become a set of evolution equations,
\begin{align}
\label{eq:EvMetric}
(\partial_{t}-\mathcal{L}_{\beta})g_{ij} &= -2NK_{ij},\\
\label{eq:EvK}
(\partial_{t}-\mathcal{L}_{\beta})K_{ij} &= N\left(R_{ij}-2K_{ik}{K^k}_j+KK_{ij}\right)-
\nabla_i\nabla_jN,
\end{align}
and a set of constraint equations,
\begin{align}
\label{eq:Ham}
\SRicciS + \TrExCurv^2 - \ExCurv_{ij}\ExCurv^{ij} & = 0,\\
\label{eq:Mom}
\nabla_j\left(\ExCurv^{ij}-\SMetric^{ij}\TrExCurv\right) & = 0.
\end{align}
In the above, $\mathcal{L}$ is the Lie derivative, $\nabla_i$ is the covariant 
derivative compatible with $\SMetric_{ij}$, 
$\SRicciS=\SMetric^{ij}\SRicci_{ij}$ is the trace of the Ricci tensor 
$\SRicci_{ij}$ of $\SMetric_{ij}$, and $K=\SMetric^{ij}\ExCurv_{ij}$ is the 
trace of the extrinsic curvature $\ExCurv_{ij}$ of $\Sigma_t$. 

Next, a conformal decomposition of various quantities is introduced. 
The conformal metric 
$\CMetric_{ij}$ and conformal factor $\CF$ are given by 
\begin{equation}
\label{eq:SMetric-ID}
g_{ij}=\CF^4\CMetric_{ij},
\end{equation}
the time derivative of the conformal metric is denoted by
\begin{equation}
\tilde{u}_{ij}=\partial_{t}\CMetric_{ij},
\end{equation}
and satisfies $\tilde{u}_{ij}\tilde{g}^{ij}=0$,
while the conformal lapse is given by $\tilde{N}=\CF^{-6}N$.
Equations~\eqref{eq:Ham},~\eqref{eq:Mom}, and the trace 
of~\eqref{eq:EvK} can then be written as 
\begin{align}
\label{eq:XCTS-Ham}
\tilde{\nabla^2}\psi-\frac{1}{8}\psi\tilde{R}-\frac{1}{12}\psi^5K^2+
\frac{1}{8}\psi^{-7}\tilde{A}_{ij}\tilde{A}^{ij} = 0,\\
\label{eq:XCTS-Mom}
\tilde{\nabla}_j\left(\frac{1}{2\tilde{N}}%
\left(\mathbb{L}\beta\right)^{ij}\right)-\tilde{\nabla}_j
\left(\frac{1}{2\tilde{N}}\tilde{u}^{ij}\right)%
-\frac{2}{3}\psi^6\tilde{\nabla}^iK = 0,\\
\label{eq:XCTS-EvK}
\tilde{\nabla}^2\left(\tilde{N}\psi^7\right)%
-\left(\tilde{N}\psi^7\right)\left(\frac{1}{8}\tilde{R}+
\frac{5}{12}\psi^4K^2+\frac{7}{8}\psi^{-8}\tilde{A}_{ij}\tilde{A}^{ij}\right)%
\notag\\
= -\psi^5\left(\partial_tK-\beta^k\partial_kK\right).
\end{align}
In the above, $\tilde{\nabla}_i$ is the covariant derivative compatible with 
$\tilde{g}_{ij}$, $\tilde{R}=\tilde{g}^{ij}\tilde{R}_{ij}$ is the trace of the 
Ricci tensor $\tilde{R}_{ij}$ of $\tilde{g}_{ij}$, $\tilde{\mathbb{L}}$ 
is the longitudinal operator,
\begin{equation}
\left(\tilde{\mathbb{L}}\beta\right)^{ij}=%
\tilde{\nabla}^i\beta^j+\tilde{\nabla}^j\beta^i-
\frac{2}{3}\tilde{g}^{ij}\tilde{\nabla}_k\beta^k,
\end{equation}
and $\tilde{A}^{ij}$ is
\begin{equation}
\tilde{A}^{ij}=\frac{1}{2\tilde{N}}%
\left(\left(\tilde{\mathbb{L}}\beta\right)^{ij}-
\tilde{u}^{ij}\right),
\end{equation}
which is related to $K_{ij}$ by
\begin{equation}
\label{eq:K-ID}
K_{ij}=\psi^{-10}\tilde{A}_{ij}+\frac{1}{3}g_{ij}K.
\end{equation}
For given $\tilde{g}_{ij}$, $\tilde{u}_{ij}$, $K$, and $\partial_tK$, Eqs.
~\eqref{eq:XCTS-Ham},~\eqref{eq:XCTS-Mom}, and~\eqref{eq:XCTS-EvK} are a 
coupled set of elliptic equations that can be solved for $\psi$, $\tilde{N}$, 
and $\beta^i$. From these solutions, the physical initial data $g_{ij}$ and 
$K_{ij}$ are obtained from~\eqref{eq:SMetric-ID} and~\eqref{eq:K-ID}, 
respectively.

\begin{figure}
\includegraphics[scale=0.5]{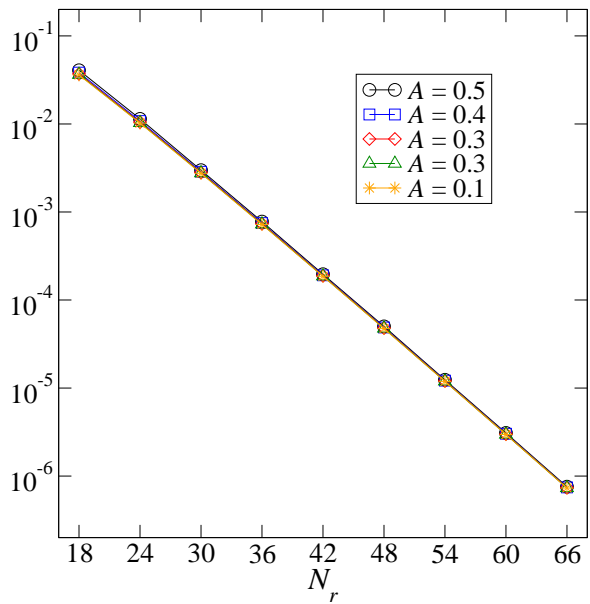}
\caption{
\label{fig:ConstraintsID_Total}
Convergence of the elliptic solver for different amplitudes $A$. 
Plotted is the square-sum of the Hamiltonian and momentum constraints, 
Eqs.~(\ref{eq:Ham}) and (\ref{eq:Mom}), as a function of numerical resolution, 
measured here by the number of radial basis functions in the spherical shell 
containing the gravitational waves.
}
\end{figure}

To construct initial data describing a Kerr black hole initially in 
equilibrium, together with an ingoing pulse of gravitational waves, we make 
the following choices for the free data,
\begin{align}
\label{eq:ConformalMetric}
\tilde{g}_{ij} &= g^{\text{KS}}_{ij}+Ah_{ij},\\
\tilde{u}_{ij} &= A\partial_th_{ij}-\frac{1}{3}\tilde{g}_{ij}\tilde{g}^{kl}
A\partial_th_{kl},\\
K &= K^{KS},\\
\partial_tK &= 0.
\end{align}
In the above, $g^{\text{{KS}}}_{ij}$ and $K^{\text{KS}}$ are the spatial metric and the trace of the extrinsic curvature in Kerr-Schild coordinates, 
with mass parameter $M_{\text{KS}}=1$ and spin parameter 
$a_{\text{KS}}=0.7M_{\text{KS}}$ along the $z$-direction. 
The pulse of gravitational waves is denoted by $h_{ij}$, and is chosen to be 
an ingoing, even parity, $m=2$, linearized quadrupole wave in a {\em flat}
background as given by Teukolsky~\cite{Teukolsky1982} 
(see~\cite{Rinne2008c} for the solution for all multipoles).
The explicit expression for the spacetime metric of the waves in 
spherical coordinates is
\begin{align}
h_{ij}dx^idx^j &= \left(R_1\sin^2\theta\cos2\phi\right)dr^2\notag\\
     &+ 2R_2\sin\theta\cos\theta\cos2\phi rdrd\theta\notag\\
     &- 2R_2\sin\theta\sin2\phi r\sin\theta drd\phi\notag\\
     &+ \left[R_3\left(1+\cos^2\theta\right)%
               \cos2\phi-R_1\cos2\phi\right]r^2d^2\theta\notag\\
     &+ \left[2\left(R_1-2R_3\right)\cos\theta\sin2\phi\right]r^2\sin\theta%
        d\theta d\phi\notag\\
     &+ \left[R_3\left(1+\cos^2\theta\right)\cos2\phi%
              +R_1\cos^2\theta\cos2\phi\right]\notag\\
     &\hspace{40 mm} \times r^2\sin^2\theta d^2\phi,
\end{align}
where the radial functions are
\begin{align}
R_1 &= 3\left[\frac{F^{(2)}}{r^3}+\frac{3F^{(1)}}{r^4}+\frac{3F}{r^5}\right],\\
R_2 &= - \left[\frac{F^{(3)}}{r^2}+\frac{3F^{(2)}}{r^3}+\frac{6F^{(1)}}{r^4}
         +\frac{6F}{r^5}\right],
\end{align}
\begin{align}
R_3 &= \frac{1}{4}\left[\frac{F^{(4)}}{r}+\frac{2F^{(3)}}{r^2}%
             +\frac{9F^{(2)}}{r^3}+\frac{21F^{(1)}}{r^4}%
             +\frac{21F}{r^5}\right],
\end{align}
and the shape of the waves is determined by
\begin{align}
F &= F(t+r) = F(x) = e^{-(x-x_0)^2/w^2},\\
F^{(n)} &\equiv \left[\frac{d^nF(x)}{dx^n}\right]_{x=t+r}.
\end{align}
We choose $F$ to be a Gaussian of width $w/M_{\text{KS}}=1.25$, 
at initial radius $x_0/M_{\text{KS}}=15$. 
The constant $A$ in Eq.~\eqref{eq:ConformalMetric} is the 
amplitude of the waves. We use the values $A=0.1$, 
0.2, 0.3, 0.4, and 0.5, each resulting in a separate initial data set.

\begin{figure}
\includegraphics[scale=0.5]{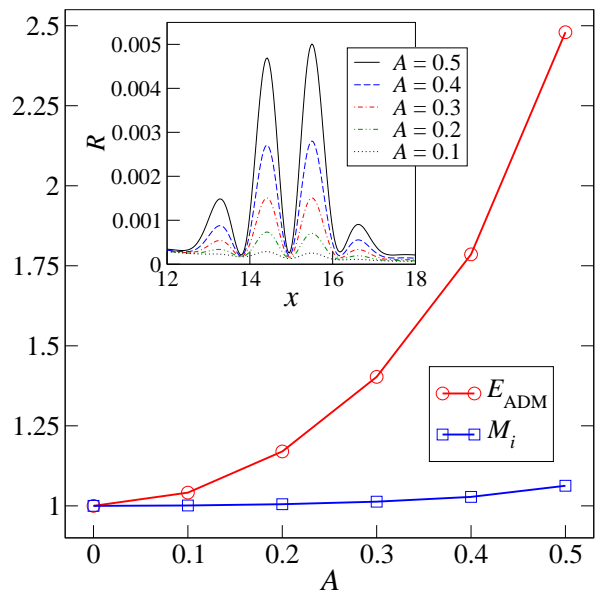}
\caption{
\label{fig:EnergyMassRicciVsAmp}
ADM energy $E_{\text{ADM}}$ and Christodoulou mass $M_i$ of the
initial data sets, versus the gravitational wave amplitude $A$. The inset
shows the Ricci scalar $R$ along the $x$-axis. All quantities are given
in units of the mass of the background Kerr-Schild metric.
}
\end{figure}

Equations~\eqref{eq:XCTS-Ham},~\eqref{eq:XCTS-Mom}, and~\eqref{eq:XCTS-EvK} are 
solved with the pseudospectral elliptic solver described in~\cite{Pfeiffer2003}.
The domain decomposition used in the elliptic solver consists of three spherical
shells with boundaries at radii $r/M_{\text{KS}}=1.5$, 12, 18, and $10^9$, 
so that the middle shell is centered on the initial location of the 
gravitational wave pulse. The inner boundary lies inside the apparent 
horizon and Dirichlet boundary conditions appropriate for the 
Kerr black hole are imposed. 
It should be noted that these boundary conditions are only 
strictly appropriate in the limit of small $A$ and large $x_0$, when the 
initial data corresponds to an ingoing pulse of linearized gravitational 
waves on an asymptotically flat background, with a Kerr black hole at the 
origin. As $A$ is increased and $x_0$ is reduced, we expect this property 
to remain qualitatively true, although these boundary conditions become 
physically less well motivated. Nonetheless, we show below by 
explicit evolution that most of the energy in the pulse moves inward and 
increases the black hole mass.
 
At the lowest resolution, the number of 
radial basis functions in each shell is (from inner to outer) $N_r=9$, 18, 
and 9, and the number of angular basis functions in each shell is $L=5$. At the 
highest resolution, the number of radial basis functions in each shell is 
(from inner to outer) $N_r=41$, 66, and 41, and the number of angular basis 
functions in each shell is $L=21$. 
Figure~\ref{fig:ConstraintsID_Total} shows 
the convergence of the elliptic solver. The expected exponential convergence 
is clearly visible.  Curves for each $A$ lie very nearly on top of 
each other, indicating that convergence is independent of the amplitude of the 
waves.
We evolve the initial data sets computed at the highest resolution of the 
elliptic solver.

We locate the apparent horizon (the outermost marginally outer trapped 
surface defined in Sec.~\ref{sec:MOTS_Intro}) in each initial data set 
using the 
pseudospectral flow method of Gundlach~\cite{Gundlach1998} (explained 
briefly in Sec.~\ref{sec:MOTS_Finders}), and compute the black hole's initial 
quasilocal angular momentum $\Spin_i$ and Christodoulou mass $M_i$ 
(the subscript ``$i$'' denotes initial values). 
The quasilocal angular momentum $\Spin$ is defined in 
Eq.~\eqref{eq:angularmom}, which we calculate with approximate 
Killing vectors~\cite{Lovelace2008} (see also~\cite{Cook2007}).
The Christodoulou mass $M$ is given by
\begin{equation}\label{eq:ChristodoulouMass}
\HorizonMass=\sqrt{{M_\text{H}}^2+\frac{\Spin^2}{4{M_\text{H}}^2}},
\end{equation}
where $M_\text{H}=\sqrt{A_\text{H}/16\pi}$ is the Hawking or irreducible 
mass~\cite{Hawking1968}, with $A_\text{H}$ being the area of the marginally 
outer trapped surface of interest. The 
main panel of Fig.~\ref{fig:EnergyMassRicciVsAmp} shows $M$ and the 
Arnowitt-Deser-Misner (ADM) 
energy $E_{\text{ADM}}$, as a function of the amplitude $A$ of each initial 
data set. The difference between $E_{\text{ADM}}$ and $M$ is a measure of 
the energy contained in the ingoing gravitational waves. For $A\gtrsim0.4$, 
this energy is comparable to or greater than $M$, so the black hole will 
become strongly distorted in the subsequent evolution. 
The inset of Fig.~\ref{fig:EnergyMassRicciVsAmp} shows the Ricci scalar $R$ 
of $g_{ij}$ along the $x$-axis at the initial location of the gravitational 
wave pulse. The sharp features of $R$ necessitate the use of the higher 
$N_r$ as labeled in Fig.~\ref{fig:ConstraintsID_Total}.

\section{Evolutions}
\label{sec:Ev}

Each of the initial data sets are evolved with the Spectral Einstein Code 
({\tt SpEC}) described in~\cite{Scheel2006,SpECwebsite}.  
This code solves 
a first-order representation~\cite{Lindblom2006} of the generalized
harmonic system~\cite{Friedrich1985,Garfinkle2002,Pretorius2005c}.
The gauge freedom in the generalized harmonic system is fixed
via a freely specifiable gauge source function $H_{\mu}$ that satisfies
\begin{equation}
  H_{\mu}(t,x) = g_{\mu\nu}\nabla_{\lambda}\nabla^{\lambda}x^{\nu}
               = -\Gamma_{\mu},
\end{equation}
where $\Gamma_{\mu}=g^{\nu\lambda}\Gamma_{\mu\nu\lambda}$ is the trace of 
the Christoffel symbol. In 3+1 form, the above expression gives evolution
equations for $N$ and $\beta^i$~\cite{Lindblom2006},
\begin{align}
  \partial_t N - \beta^i\partial_i N &= -N\left(H_t - \beta^iH_i + NK\right),\\
  \partial_t \beta^i - \beta^k\partial_k \beta^i &= 
    Ng^{ij}\left[N\left(H_j + g^{kl}\Gamma_{jkl}\right) - \partial_j N\right],
\end{align}
so there is no loss of generality in specifying $H_{\mu}$ instead of $N$ and 
$\beta^i$, as is more commonly done. For our evolutions, $H_{\mu}$ is 
held fixed at its initial value.

The decomposition of the computational domain consists of eight concentric 
spherical shells surrounding the black hole. The inner boundary of the 
domain is at $r/M_{\text{KS}}=1.55$, inside the apparent horizon of the 
initial black hole, while the outer boundary is at $r/M_{\text{KS}}=50$. 
The outer 
boundary conditions~\cite{Lindblom2006,Rinne2006,Rinne2007} are designed to
prevent the influx of unphysical constraint violations~\cite{Stewart1998,%
FriedrichNagy1999,Bardeen2002,Szilagyi2002,%
Calabrese2003,Szilagyi2003,Kidder2005} and undesired incoming gravitational
radiation~\cite{Buchman2006,Buchman2007}, while allowing the outgoing
gravitational radiation to pass freely through the boundary. Interdomain
boundary conditions are enforced with a penalty method
~\cite{Gottlieb2001,Hesthaven2000}. The evolutions were run on up to three 
different resolutions -- low, medium, and high. For the low resolution, 
the number of radial basis functions in each shell is 
$N_r=23$, and the number of angular basis functions in each shell is 
$L=15$. For the high resolution, $N_r=33$ and $L=21$ in each shell.

\begin{figure}
\includegraphics[scale=0.5]{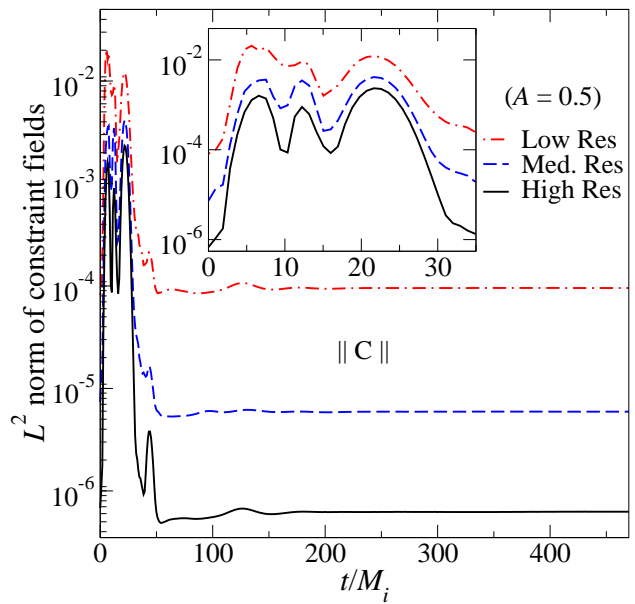}
\caption{
\label{fig:ConstraintsEv}
Constraint violations for the evolution with $A=0.5$. Plotted is
the $L^2$ norm of all constraints, normalized by the $L^2$ norm of the spatial
gradients of all dynamical fields.
}
\end{figure}

We will be mainly interested in the case where the gravitational waves 
have an amplitude $A=0.5$. As a measure of the accuracy of this evolution, 
the constraints
of the first-order generalized harmonic system are plotted in
Fig.~\ref{fig:ConstraintsEv}. 
Plotted is the $L^2$ norm of all constraint fields, normalized by the 
$L^2$ norm of the spatial gradients of the dynamical fields 
(see Eq. (71) of~\cite{Lindblom2006}).
The $L^2$ norms are taken over the entire computational volume. 
The constraints increase at first, as the black hole is 
distorted by the gravitational waves. 
As the black hole settles down to 
equilibrium, the constraints decay and level off. The results presented in the 
following sections use data from the high resolution runs only.

\section{Marginally Trapped Surfaces}
\label{sec:MTS}

\subsection{Basic Definitions and Concepts}
\label{sec:MOTS_Intro}
Let $\mathcal{S}$ be a closed, orientable spacelike 2-surface in $\Sigma_t$. 
There are two linearly independent and future-directed outgoing and ingoing 
null vectors $l^{\mu}$ and $k^{\mu}$ normal to $\mathcal{S}$. We write these 
vectors in terms of the future-directed timelike unit normal $n^{\mu}$ to 
$\Sigma_t$ and the outward-directed spacelike unit normal 
$s^{\mu}$ to $\mathcal{S}$ as
\begin{align}
\label{eq:nullnormals}
l^{\mu}=\frac{1}{\sqrt{2}}\left(n^{\mu}+s^{\mu}\right)%
\hspace{2 mm}\text{and}\hspace{2 mm}%
k^{\mu}=\frac{1}{\sqrt{2}}\left(n^{\mu}-s^{\mu}\right),
\end{align}
normalized so that $g_{\mu\nu}l^{\mu}k^{\nu}=-1$. Then the induced metric
$\bar{q}_{\mu\nu}$ on $\mathcal{S}$ is
\begin{align}
\bar{q}_{\mu\nu} &= g_{\mu\nu} + l_{\mu}k_{\nu} + l_{\nu}k_{\mu},\\
                   &= g_{\mu\nu} + n_{\mu}n_{\nu} - s_{\mu}s_{\nu}.
\end{align} 

The extrinsic curvatures of $\mathcal{S}$ as embedded 
in the full four-dimensional spacetime are
\begin{align}
\label{eq:NullExtrinsicCurvatures}
\bar{K}^{(l)}_{\mu\nu} = \bar{q}^{\lambda}_{\mu}\bar{q}^{\rho}_{\nu}%
                         \nabla_\lambda l_\rho%
\hspace{2 mm}\text{and}\hspace{2 mm}%
\bar{K}^{(k)}_{\mu\nu} = \bar{q}^{\lambda}_{\mu}\bar{q}^{\rho}_{\nu}%
                         \nabla_\lambda k_\rho.
\end{align}
The null vectors $l^\mu$ and $k^\mu$ are tangent to a congruence of 
outgoing and ingoing null geodesics, respectively. The traces of 
the extrinsic curvatures give the congruences' $\textit{expansions}$
\begin{align}
\label{eq:nullexpansions}
\theta_{(l)} = \bar{q}^{\mu\nu}\nabla_{\mu}l_{\nu}%
\hspace{2 mm}\text{and}\hspace{2 mm}%
\theta_{(k)} = \bar{q}^{\mu\nu}\nabla_{\mu}k_{\nu},
\end{align}
and the $\textit{shears}$ are the trace-free parts,
\begin{align}
\label{eq:nullshears}
\sigma^{(l)}_{\mu\nu} &= \bar{q}^{\lambda}_{\mu}\bar{q}^{\rho}_{\nu}\nabla_{\lambda}l_{\rho} - \frac{1}{2}\bar{q}_{\mu\nu}\theta_{(l)}%
\hspace{2 mm}\text{and}\\
\sigma^{(k)}_{\mu\nu} &= \bar{q}^{\lambda}_{\mu}\bar{q}^{\rho}_{\nu}\nabla_{\lambda}k_{\rho} - \frac{1}{2}\bar{q}_{\mu\nu}\theta_{(k)}.
\end{align}

The geometrical interpretation of the expansion is the fractional rate of 
change of the congruence's cross-sectional area~\cite{PoissonToolkit}.
We will mainly be interested in 2-surfaces $\mathcal{S}$ on which 
$\theta_{(l)}=0$, called $\textit{marginally outer trapped surfaces}$ 
(MOTSs) following the terminology in~\cite{Schnetter2006}. 
If $\theta_{(l)}<0$ on $\mathcal{S}$, then outgoing null normals 
will be converging towards each other, as one expects to happen inside 
a black hole. If $\theta_{(l)}>0$ the situation is reversed, so the 
condition $\theta_{(l)}=0$ provides a reasonable quasilocal prescription 
for identifying the surface of a black hole. In practice, an MOTS will 
generally lie inside the event horizon, unless the black hole is 
stationary. The outermost MOTS is called the $\textit{apparent horizon}$, 
and is used to represent the surface of a black hole in numerical simulations.
In the next subsection, we briefly describe how we locate MOTSs.

\subsection{MOTS Finders}
\label{sec:MOTS_Finders}

We use two different algorithms to locate MOTSs in $\Sigma_t$.  
Both algorithms expand an MOTS ``height function''
in spherical harmonics
\begin{equation}\label{eq:MOTS}
r_{\rm MOTS}(\theta, \phi)=\sum_{l=0}^{L_{\rm MOTS}}\sum_{m=-l}^lA_{lm}Y_{lm}(\theta, \phi).
\end{equation}
Our standard algorithm is the pseudospectral fast flow method developed by
Gundlach~\cite{Gundlach1998}, which we use during the evolution.  
This method utilizes the fact that the
MOTS condition $\theta_{(l)}=0$ results in an elliptic equation for
$r_{\rm MOTS}(\theta,\phi)$. The elliptic equation is solved using a
fixed-point iteration with the flat-space Laplacian on $S^2$ on the
left-hand side, which is computationally inexpensive to invert given
the expansion Eq.~(\ref{eq:MOTS}).  The fixed-point iteration is
coupled to parameterized modifications which allow for tuning of the method
to achieve fast, but still reasonably robust convergence.
In Gundlach's nomenclature, we use the N flow method, and have found the 
parameters $\alpha=1$ and $\beta=0.5$ satisfactory (see~\cite{Gundlach1998} 
for definitions).

Gundlach's algorithm (as well as MOTS finders based on flow methods in
general~\cite{nakamura84,tod91}) incorporates a sign assumption on the 
surfaces near the MOTS, namely that $\theta_{(l)}$ is 
{\em positive} for a 
surface which lies somewhat {\em outside} of the MOTS.  This assumption is
satisfied for the apparent horizon.
However, this sign assumption is not satisfied for some inner MOTSs in 
$\Sigma_t$ that we discover below. Therefore, these inner MOTSs are 
unstable fixed-points for Gundlach's algorithm, so that this algorithm cannot 
locate these MOTSs.

To find these inner MOTSs, we employ an older algorithm that is based on a
minimization technique~\cite{baumgarte_etal96,Pfeiffer2000,Pfeiffer2002a}: 
The coefficients $A_{lm}$ in Eq.~(\ref{eq:MOTS}) 
are determined by minimizing the functional
\begin{equation}
\Theta\equiv\int_{\mathcal{S}} \theta_{(l)}^2\,\sqrt{\bar{q}}d^2 x
\end{equation}
where the surface integral is over the current trial surface with 
area element $\sqrt{\bar{q}}$.  This technique
is insensitive to the sign assumption in Gundlach's method.  However, it is
much slower, especially for large $L_{\rm MOTS}$.

When multiple MOTSs are present in $\Sigma_t$, the choice of an
initial surface determines the final surface the MOTS finder converges to.
Therefore, both MOTS finders require judicious choices of these initial
surfaces.  We typically track MOTSs from time step to time step, and use the 
MOTS at the  previous time step as an initial guess for the MOTS finder at the 
current time.

\section{Marginally Trapped Tubes}
\label{sec:MTT}

\subsection{Basic Definitions and Concepts}
\label{sec:MTT_Intro}
During an evolution, the MOTSs found
at successive times foliate a world tube, or a
$\textit{marginally trapped tube}$ (MTT).
The type of MTT that is foliated by a series of MOTSs depends on the 
physical situation. A null MTT is an $\textit{isolated %
horizon}$~\cite{Ashtekar2000_Generic,Ashtekar2001,Ashtekar1999,%
Ashtekar2000,Ashtekar2000_Ham} if $-R_{\mu\nu}l^{\nu}$ is future causal, 
and certain quantities are time independent on it. An isolated horizon 
describes a black hole in equilibrium. On the other hand, a 
$\textit{dynamical horizon}$ describes a black hole that is 
absorbing matter or gravitational 
radiation~\cite{Ashtekar2002,Ashtekar2003}, and is physically the most 
relevant. A dynamical horizon is a spacelike MTT foliated by MOTSs on 
which $\theta_{(k)}<0$, called  
$\textit{future marginally outer trapped surfaces}$. 
For a given slicing of spacetime by spatial hypersurfaces 
$\Sigma_t$, the foliation of a dynamical horizon by future marginally 
outer trapped surfaces 
on $\Sigma_t$ is unique~\cite{Ashtekar2005}. Since the location of a 
MOTS is a property of $\Sigma_t$, different spacetime slicings will in 
general give different MTTs. Also, a timelike MTT is called a 
$\textit{timelike membrane}$~\cite{Ashtekar-Krishnan:2004}. 
Since causal curves can traverse it in both inward and outward 
directions, it cannot represent the surface of a black hole.

An additional characterization of MTTs is based on trapping 
horizons~\cite{Hayward1994}. A $\textit{future outer trapping horizon}$ 
is an MTT foliated by MOTSs that have $\theta_{(k)}<0$ and 
${\cal L}_k\theta_{(l)}<0$ for some scaling of $l^{\mu}$ and $k^{\mu}$. 
Such an MOTS is called a $\textit{future outer trapping surface}$.
If the null energy condition holds, a future outer trapping horizon 
is either completely null or 
completely timelike. It was shown in~\cite{BoothFairhurst2007} that if 
${\cal L}_k\theta_{(l)}\neq0$ for at least one point on these future 
outer trapping surfaces, then 
the future outer trapping horizon is spacelike, or a dynamical horizon, 
in a neighborhood of the future outer trapping surfaces. 
Otherwise the future outer trapping horizon is null.

Interestingly, an MTT may not fall into either of the categories described 
above, but can have sections of mixed signatures as demonstrated in the 
head-on collision of two black holes~\cite{Schnetter2006}.
At merger, a common apparent horizon appears in $\Sigma_t$ that surrounds 
the MOTSs of the individual black holes. This common horizon then 
bifurcates into outer and inner common horizons. The outer common horizon 
grows in area and is spacelike. However, the inner common
horizon decreases in area and foliates an MTT that is briefly partly
spacelike and partly timelike, before becoming a timelike membrane later on.

\subsection{Multiple MTTs}
\label{sec:MultipleMTT}
\begin{figure}
\includegraphics[scale=0.5]{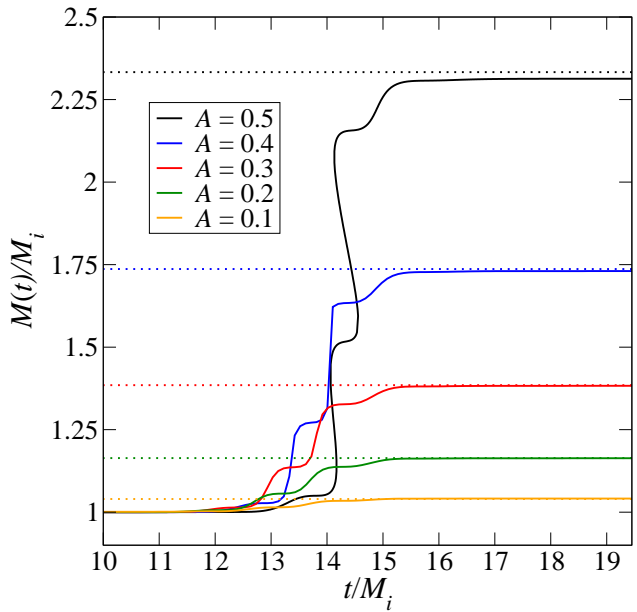}
\caption{
\label{fig:MOTSMass}
The solid curves are the Christodoulou masses $M(t)$ divided by their
initial values $M_i$ for the five evolutions with different amplitudes
$A=0.1$, 0.2, 0.3, 0.4, and 0.5 for the ingoing pulse of gravitational waves.
The horizontal dotted lines denote the ADM energy of each data set,
$E_{\text{ADM}}/M_i$.
}
\end{figure}

\begin{figure}
\includegraphics[scale=0.5]{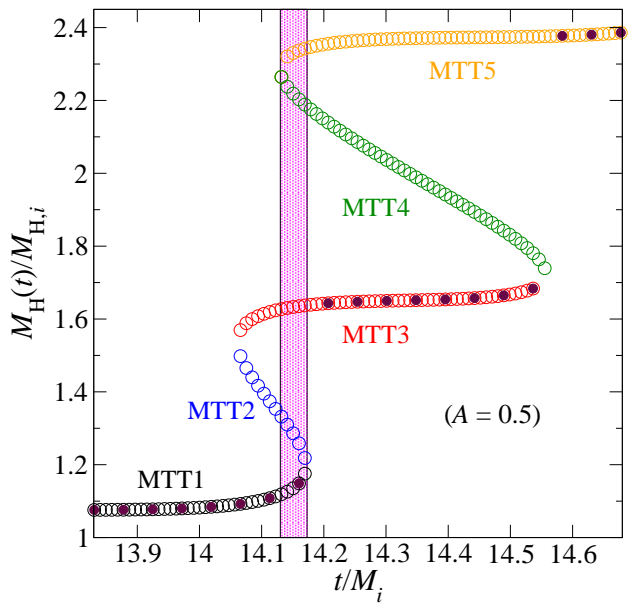}
\caption{
\label{fig:MOTSMass5}
Irreducible mass $M_\text{H}$ divided by its initial value 
$M_{\text{H},i}$ for the evolution with $A=0.5$. The solid 
circles are the values of $M_\text{H}$ for MOTSs found during the evolution.
The completed curve is traced out by open circles. The vertical shaded region
indicates when five MOTSs exist at the same time.
}
\end{figure}

We now discuss the MOTSs that occur during the five 
evolutions of the distorted black hole,    
with  amplitude $A=0.1$, 0.2, 0.3, 0.4, or 0.5 
for the ingoing gravitational wave pulse.  
The MOTSs we find are indicated in Fig.~\ref{fig:MOTSMass} by their 
Christodoulou masses $M$.
Early in each simulation, $M$ is approximately 
constant, and begins to increase when the gravitational wave hits 
the black hole around $t\approx 12M_i$. 
The effect is more pronounced for larger $A$. 
The horizontal dotted lines in 
Fig.~\ref{fig:MOTSMass} indicate the ADM energy of the initial data. 
Although we do not explicitly calculate the energy carried away by 
gravitational waves, we can still see that the final 
{\em Christodoulou} mass is close to $E_{\rm ADM}$, 
indicating that the energy in the gravitational wave pulse predominantly falls
into the black hole, and only a small fraction of this energy
propagates to null infinity. 
Even for the highest amplitude case of $A=0.5$, the final value of 
$M$ is about $99.1\%$ of the ADM energy.
These results are as expected.  However, for both $A=0.4$ and $A=0.5$, 
a very interesting new
feature arises: {\em multiple} concentric MOTSs are present at the 
{\em same} coordinate time.

The evolution with $A=0.5$ shows the multiple MOTSs more distinctly, 
hence we will focus on it in the remainder of this paper. 
Figure~\ref{fig:MOTSMass5} presents a closer look at the 
{\em irreducible} masses $M_\text{H}$ for this case. 
Locating all MOTSs shown in Fig.~\ref{fig:MOTSMass5} requires considerable
care.  The starting point was the output of the MOTS finder that was run
during the evolution, using Gundlach's fast flow
algorithm~\cite{Gundlach1998}.  Because of the computational expense involved,
the MOTS finder was not run very frequently, resulting in the
solid circles in Fig.~\ref{fig:MOTSMass5}.
The MOTS at the previous time was used
as the initial guess for the current time, resulting in a series of MOTSs
which is as continuous as possible. The curve traced out by these
points has sharp jumps, which was the first indication of the presence of
multiple MOTSs at these times. Then to find the remainder of MTT3
and MTT5, an MOTS corresponding to one of these solid circles on MTT3 or MTT5
was used as an initial guess and the MOTS finder was also run more frequently.
At this stage, we had completely traced out MTT1, MTT3, and MTT5.
Next we found MTT2 and MTT4 to be unstable fixed points for Gundlach's
algorithm, so it was necessary to use our older MOTS finder based on a
minimization technique~\cite{baumgarte_etal96,Pfeiffer2000,Pfeiffer2002a}
to find these MTTs.
As an initial guess for finding an MOTS on MTT2
for instance, a sphere with radius equal to the average radii of
MTT1 and MTT3 sufficed.
Once an MOTS on MTT2 was located, it was used as an initial guess for the
MOTS finder to locate the MOTSs on neighboring time slices 
(both later and earlier).  The same procedure was used to locate MTT4.

After finding all the MTTs in Fig.~\ref{fig:MOTSMass5}, a clearer picture 
of their structures in relation to each other emerged. MTT1 corresponds to 
the surface of the initial black hole. Shortly after
$t=14M_i$, a new MOTS with $M_{\text{H}}/M_{\text{H},i}\approx1.525$ 
appears and bifurcates 
into two MTTs. $M_\text{H}$ decreases along MTT2, which promptly annihilates 
with MTT1, while MTT3 persists slightly longer. A similar process then takes 
place again, and MTT5 is left over as the surface of the
final black hole, with $M_{\text{H}}$ more than double its initial value. 
The vertical shaded region indicates the time interval when five MTTs exist
simultaneously. Notice that $M_{\text{H}}$ of the apparent horizon
jumps discontinuously in time from the curve of MTT1 to MTT3, and then to 
MTT5. This indicates that the apparent horizon itself is discontinuous across
these times.

\begin{figure}
\includegraphics[scale=0.5]{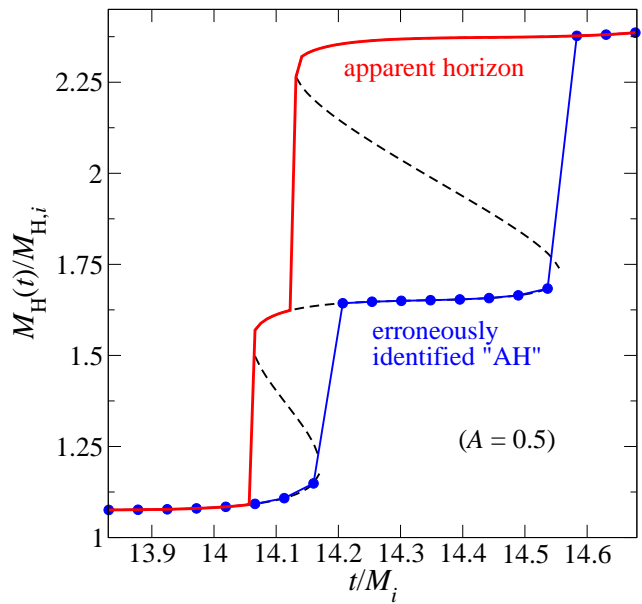}
\caption{
\label{fig:naiveAH}
The solid red line denotes the apparent horizon for the evolution with 
$A=0.5$.  The solid blue circles denote an erroneous
``apparent horizon,'' which is found when the apparent horizon
finder is run during the evolution in larger time intervals. The black dashed
lines denotes all five MTTs as shown in Fig.~\ref{fig:MOTSMass5}. }
\end{figure}

  The apparent horizon is the outermost MOTS, and when 
  only one MOTS is present in a black hole evolution, the MOTS and
  apparent horizon are identical.  Here this is not the case, and
  Fig.~\ref{fig:naiveAH} shows the apparent horizon in relation to the
  various MTTs.  This figure also highlights another potential pitfall
  when locating MOTSs.  MOTS finders are typically run during the
  evolution fairly infrequently, using the MOTS from the last MOTS
  computation as an initial guess (to minimize computational cost).  If
  this had been done for the $A=0.5$ case shown in
  Figs.~\ref{fig:MOTSMass5} and~\ref{fig:naiveAH}, the solid blue
  circles would have been obtained.  Because the previously found MOTS
  is used as an initial guess, newly appearing MOTSs are generally
  missed.  For instance, the solid blue circles follow MTT1 until it
  disappears, instead of jumping to MTT3.  Therefore, the output of
  the ``apparent horizon finder'' (the more widely used name, but
  technically less precise than ``MOTS finder''), is sometimes {\em not}
  the apparent horizon.

\begin{figure}
\includegraphics[scale=0.5]{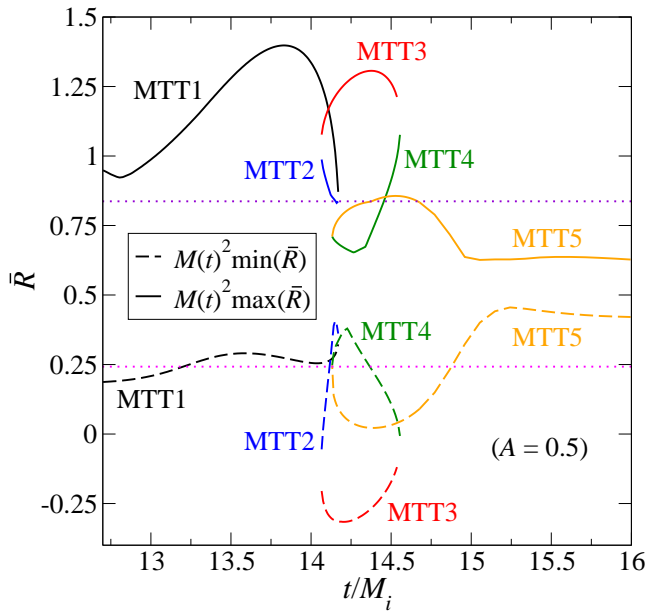}
\caption{
\label{fig:MinMaxRicci5}
Extrema of the intrinsic scalar curvature $\bar{R}$ of
MOTSs during the evolution with $A=0.5$. The horizontal dotted lines
are the values for the apparent horizon in the initial data.
Around $t=14.25M_i$, the MOTSs have regions of negative $\bar{R}$.
}
\end{figure}

A measure of the distortion of the black hole is provided by the intrinsic
scalar curvature $\bar{R}$ of the MOTSs. The extrema of $\bar{R}$ is shown in
Fig.~\ref{fig:MinMaxRicci5}, along with those of the initial apparent 
horizon. It is interesting to point out that around $t=14.25M_i$, the 
distortion caused by the gravitational waves with 
$A=0.5$ is sufficiently strong to produce regions of negative $\bar{R}$.

\subsection{Dynamical Horizons}
\label{sec:DH}
\begin{figure}
\includegraphics[scale=0.5]{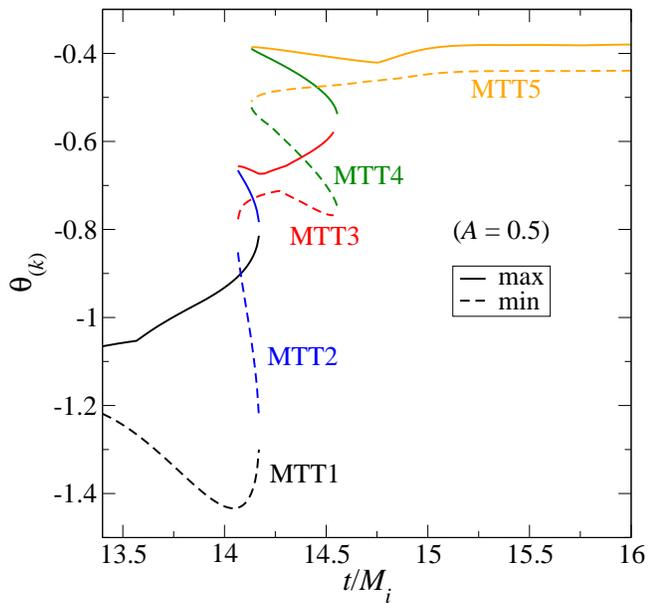}
\caption{
\label{fig:MinMaxThetaIn5}
Extrema of $\theta_{(k)}$ on each MOTS along the MTTs during the
evolution with $A=0.5$. For the time shown, $\theta_{(k)}<0$.
}
\end{figure}

\begin{figure}
\includegraphics[scale=0.5]{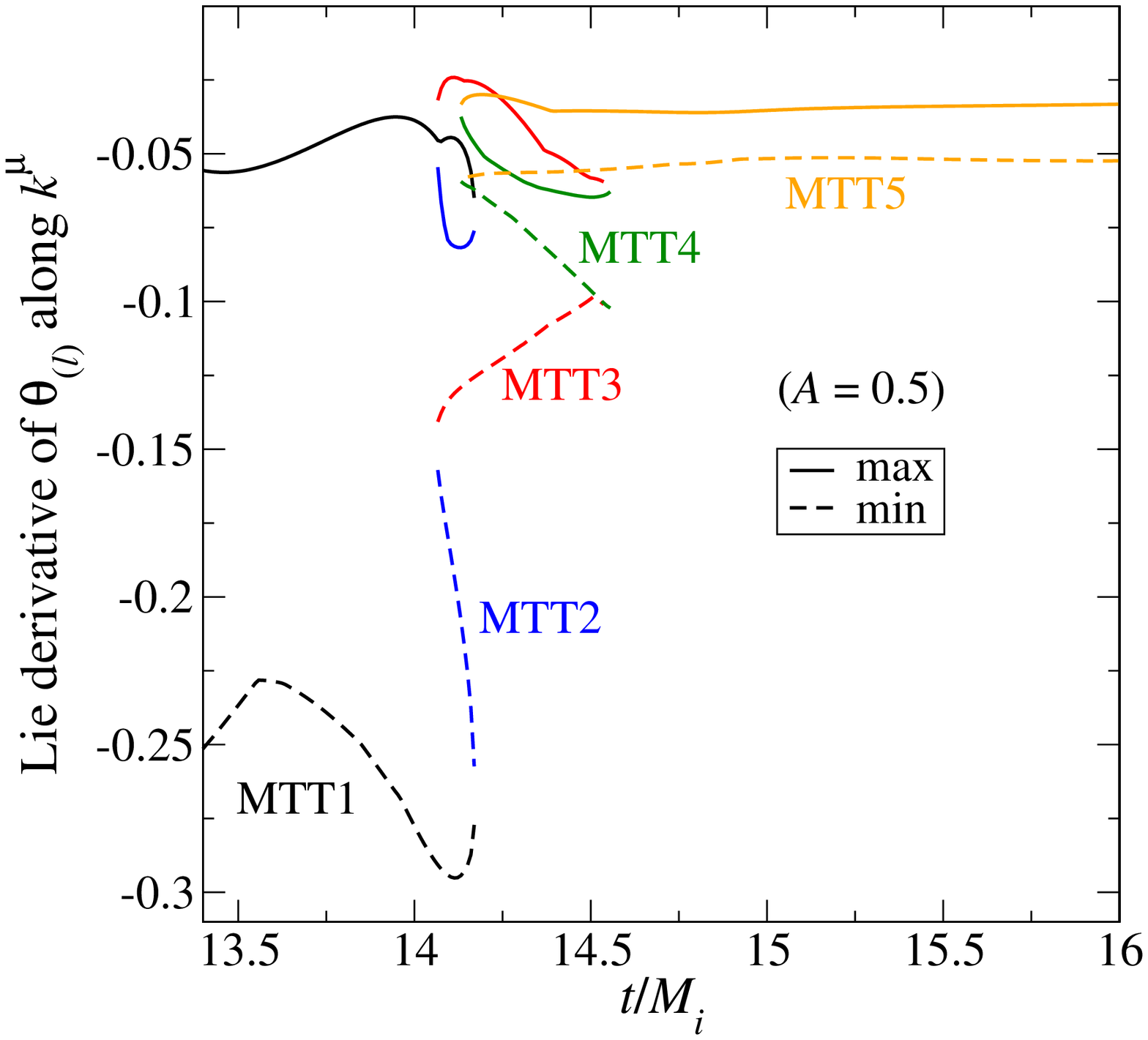}
\caption{
\label{fig:MinMaxThetaOutAlongIn5}
Extrema of ${\cal L}_k\theta_{(l)}$ on each MOTS along the MTTs
during the evolution with $A=0.5$. For the time shown,
${\cal L}_k\theta_{(l)}<0$.
}
\end{figure}

We determine the signatures of the multiple MTTs during the highly 
dynamical period. First we compute $\theta_{(k)}$ and 
$\mathcal{L}_k \theta_{(l)}$ using the null normals in 
Eq.~\eqref{eq:nullnormals}, and find that both quantities are negative. 
So our MTTs are future outer trapping horizons, which 
must be either spacelike or null, and we can 
immediately rule out the possibility of there being sections of mixed 
signatures. Figure~\ref{fig:MinMaxThetaIn5} 
shows the extrema of $\theta_{(k)}$ along each MTT. 
The quantity ${\cal L}_k\theta_{(l)}$ is evaluated from the 
expression~\cite{BoothFairhurst2007}
\begin{align}
{\cal L}_k\theta_{(l)} = -\bar{R}/2 + \omega_{\mu}\omega^{\mu} 
                         - d_{\mu}\omega^{\mu} 
                         + 8\pi T_{\mu\nu}l^{\mu}k^{\nu},
\end{align}
where 
\begin{equation}
\label{eq:omega}
\omega_{\mu}=-\bar{q}^{\nu}_{\mu}k_{\lambda}\nabla_{\nu}l^{\lambda}
\end{equation}
is the $\textit{normal fundamental form}$, and 
$d_{\mu}$ is the covariant derivative compatible with $\bar{q}_{\mu\nu}$.
Figure~\ref{fig:MinMaxThetaOutAlongIn5} shows the extrema of 
${\cal L}_k\theta_{(l)}<0$ along each MTT. 

\begin{figure}
\includegraphics[scale=0.5]{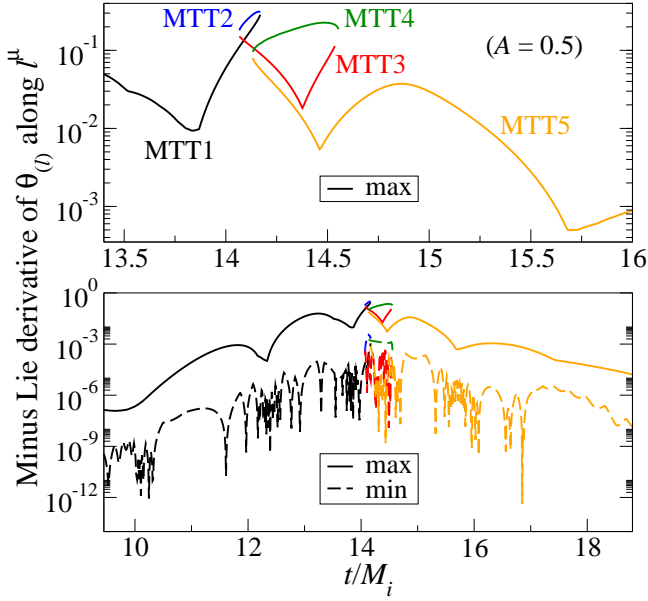}
\caption{
\label{fig:MinMaxThetaOutAlongOut5}
Extrema of $-{\cal L}_l\theta_{(l)}$ on each MOTS along the MTTs
during the evolution with $A=0.5$. Near $t=14M_i$,
${\cal L}_l\theta_{(l)}\neq0$ somewhere on $\mathcal{S}_v$.
}
\end{figure}

Next we compute ${\cal L}_l\theta_{(l)}$ to determine whether the MTTs are 
spacelike or null. We evaluate this using the null Raychaudhuri 
equation~\cite{PoissonToolkit},
\begin{equation}
\label{eq:Raychaudhuri}
{\cal L}_l\theta_{(l)} = -\sigma^{(l)\mu\nu}\sigma^{(l)}_{\mu\nu}
                         -8\pi T_{\mu\nu}l^{\mu}l^{\nu}.
\end{equation}
Figure~\ref{fig:MinMaxThetaOutAlongOut5} shows that during the times 
when there are multiple MTTs, ${\cal L}_l\theta_{(l)}\neq0$ somewhere 
on each MOTS. Thus all of the MTTs are dynamical horizons at these times. 

Here we also mention the $\textit{extremality parameter}$ $e$ of a  
MTT introduced in~\cite{BoothFairhurst:2008}. In vacuum, it is given by
\begin{align}
\label{eq:extremality}
e &= \frac{1}{4\pi}\int_{\mathcal{S}} \! \omega_\mu \omega^\mu \, \sqrt{\bar{q}}d^2 x,\\
  &= 1 + \frac{1}{4\pi}\int_{\mathcal{S}} \! \mathcal{L}_k \theta_{(l)}  \, \sqrt{\bar{q}}d^2 x,
\end{align}
where the integral is over an MOTS $\mathcal{S}$ that foliates the MTT. 
When $\mathcal{S}$ is axisymmetric, this can be regarded as the sum of 
the squares of all angular momentum multipoles.
Because a future outer trapping horizon, which is either spacelike or null, 
has $\mathcal{L}_k \theta_{(l)}<0$, it is always subextremal ($e<1$). So a 
timelike membrane foliated by future MOTSs (with $\theta_{(k)}<0$) 
must have $\mathcal{L}_k \theta_{(l)}>0$, and is superextremal ($e>1$).
Therefore, it was suggested 
in~\cite{BoothFairhurst:2008} that an MTT's transition from being spacelike 
to timelike can be detected when $e\rightarrow1$. 

\begin{figure}
\includegraphics[scale=0.5]{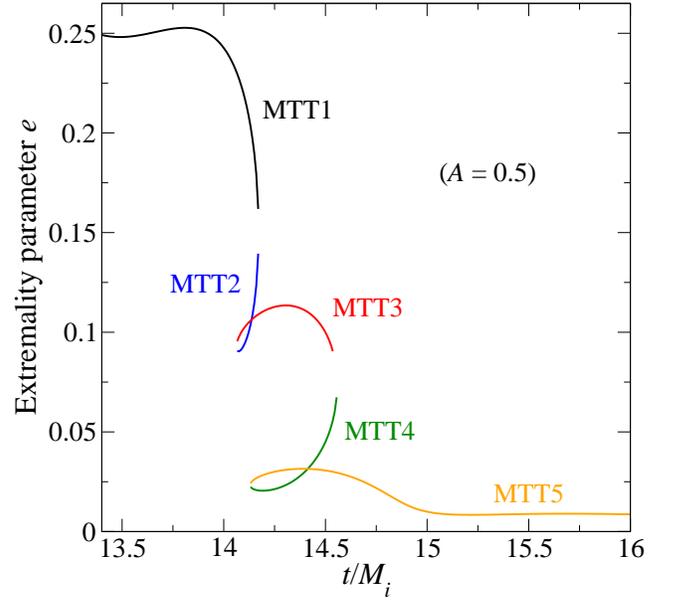}
\caption{
\label{fig:ExtremalityParam5}
Extremality parameter $e$ along the MTTs
during the evolution with $A=0.5$. For the time shown,
the MTTs are subextremal with $e<1$, indicating that the
MTTs have no timelike sections.
}
\end{figure}

Figure~\ref{fig:ExtremalityParam5} shows $e$ along each MTT, and we see 
that nowhere does $e\rightarrow1$, confirming that our MTTs do not become 
timelike. The value of $e$ shows a substantial decrease after the 
distortion has left, 
which is not due to a loss of quasilocal angular momentum 
$J$ (defined in Eq.~\eqref{eq:angularmom}), but to the large gain in 
irreducible mass $M_{\text{H}}$.
It may seem that $e$ in Fig.~\ref{fig:ExtremalityParam5} is already 
rather small to start out with, but one must recall that $e$ depends on 
the scaling of the null normals $l^\mu$ and $k^\mu$.
That is, we can define new null normals $\bar{l}^\mu=fl^\mu$ and 
$\bar{k}^\mu=k^{\mu}/f$, rescaled by some function $f$ such that the 
normalization $\bar{l}^\mu \bar{k}_\mu = -1$ is preserved. 
Then $e$ will change as
\begin{equation}
\label{ExtremalityRescale}
\bar{e} = e + \frac{1}{4\pi} \int_{\mathcal{S}} \! \left[2\omega^\mu d_\mu \text{ln} f + (d_\mu \text{ln} f) (d^\mu \text{ln} f)\right] \, \sqrt{\bar{q}}d^2 x.
\end{equation}
Nevertheless, the extremality classification of the MTTs is invariant. 

It is known that the irreducible mass $M_{\text{H}}$ of an MOTS must 
increase along a dynamical horizon~\cite{Ashtekar2003}, so at first it 
may seem surprising that MTT2 and MTT4, with decreasing $M_{\text{H}}$ 
during the evolution, are also dynamical horizons. 
However, all these MTTs can be viewed as sections of a single dynamical 
horizon $\mathcal{H}$ that weaves forwards and backwards in time. Then it 
is clear that the tangent vector to $\mathcal{H}$ along MTT2 and MTT4 points
backwards in time, so that $M_{\text{H}}$ is actually increasing along
$\mathcal{H}$ as expected. Our simple choice of holding the 
gauge source function $H_\mu$ equal to its initial value leads to 
a spacetime foliation that interweaves $\mathcal{H}$. 
This could be avoided by an alternative choice of $H_\mu$ that results in  
a single dynamical horizon that only grows in time.

The situation here resembles an example of a Tolman-Bondi spacetime 
considered in~\cite{Booth2006}, where multiple spherically symmetric 
dust shells fall into a black hole. For their chosen matter distribution, 
multiple MTTs also formed (up to three at the same time), which were either
completely spacelike, or null when the matter density vanished between
successive dust shells. In our case the role of the matter density is
replaced by the shear $\sigma^{(l)}_{\mu\nu}$ due to the gravitational 
waves. Since this is always nonvanishing somewhere on the multiple MTTs 
that form, we only have dynamical horizons. 

In~\cite{Ashtekar2005}, it was shown that for a regular dynamical horizon 
(which is achronal and also a future outer trapping horizon), 
no weakly trapped surface 
(on which $\theta_{(l)}\leq0$ and $\theta_{(k)}\leq0$) can exist 
in its past domain of dependence. This helps to explain the difficulty in 
locating MOTSs along MTT2 and MTT4 using flow methods. For example, 
consider locating an MOTS on MTT2 at $t=14.1M_i$ shown in 
Fig.~\ref{fig:MOTSMass5}. If we use a trial surface $\mathcal{S}$ located 
between the MOTSs on MTT1 and MTT2, it must have $\theta_{(l)}>0$ because it 
lies in the past domain of dependence of $\mathcal{H}$. This means that 
$\mathcal{S}$ will be moved inwards when using flow methods, away from MTT2. 
If we switch to having $\mathcal{S}$ lie between the MOTSs on MTT2 and MTT3, 
then having $\theta_{(l)}>0$ is desired. Unfortunately, now $\mathcal{S}$ 
lies in the future domain of dependence of $\mathcal{H}$, and 
we are no longer guaranteed that $\mathcal{S}$ is not a weakly trapped 
surface.

\subsection{Dynamical Horizon Flux Law}
\label{sec:DHFlux}

The growth of a black hole in full, nonlinear general relativity can be 
described by the dynamical horizon flux law of Ashtekar and 
Krishnan~\cite{Ashtekar2002,Ashtekar2003}, which relates the increase 
in area or mass along a dynamical horizon to fluxes of matter and gravitational 
energy across it. Here, we will evaluate this flux law for the dynamical 
horizon $\mathcal{H}$ that consists of the multiple MTT sections we found 
earlier, using the form given in~\cite{BoothFairhurst2007}. 

To state the dynamical horizon flux law, let us specifically consider 
the change in the irreducible mass $M_\text{H}$ along $\mathcal{H}$. 
Denote an MOTS that 
foliates $\mathcal{H}$ by $\mathcal{S}_v$, which is labeled by a foliation 
parameter $v$ that is constant on $\mathcal{S}_v$. Then choose a tangent 
vector $V^{\mu}$ to $\mathcal{H}$ that is normal to each $\mathcal{S}_v$, 
and such that 
\begin{equation}
\label{eq:foliationparam}
{\cal L}_Vv=1.
\end{equation}
This vector $V^{\mu}$ can be written as 
\begin{equation}
\label{eq:DHtangentrescaled}
V^{\mu} = \bar{B}\bar{l}^{\mu}-\bar{C}\bar{k}^{\mu}, 
\end{equation}
in terms of coefficients $\bar{B}$ and $\bar{C}$, and null normals 
$\bar{l}^{\mu}=fl^{\mu}$ and $\bar{k}^{\mu}=k^{\mu}/f$ that are rescaled 
by a function $f$ (but still having $\bar{l}^{\mu}\bar{k}_{\mu}=-1$) so that 
\begin{equation}
\label{eq:Cbar}
\bar{C} = 2\frac{d M_{\text{H}}}{d v}.
\end{equation}
The dynamical horizon flux law is then 
\begin{equation}
\label{eq:DHfluxlaw}
\frac{d M_{\text{H}}}{d v} = \int_{\mathcal{S}_v}%
  \! \left[T_{\mu\nu}\bar{l}^{\mu}\tau^{\nu}%
  + \frac{\bar{B}}{8\pi} \sigma^{(\bar{l})}_{\mu\nu}\sigma^{(\bar{l})\mu\nu}%
  + \frac{\bar{C}}{8\pi} \bar{\omega}_{\mu}\bar{\omega}^{\mu} \right] \, \sqrt{\bar{q}}d^2 x,
\end{equation}
where $\sigma^{(\bar{l})}_{\mu\nu}$ and $\bar{\omega}_{\mu}$ are given 
by Eqs.~\eqref{eq:nullshears} and~\eqref{eq:omega} but in terms of 
$\bar{l}^{\mu}$ and $\bar{k}^{\mu}$, and 
$\tau^{\mu}=\bar{B}\bar{l}^{\mu}+\bar{C}\bar{k}^{\mu}$ is the 
normal vector to $\mathcal{H}$.

The first term in Eq.~\eqref{eq:DHfluxlaw} involving $T_{\mu\nu}$ 
is the energy flux of matter across $\mathcal{S}_v$, and the second  
term involving $\sigma^{(\bar{l})}_{\mu\nu}$ is a flux of gravitational 
energy~\cite{Ashtekar2003}. The last term  
has been interpreted differently by various authors.
The normal fundamental form $\omega_\mu$ (or $\bar{\omega}_\mu$) 
enters into the definition of the quasilocal angular momentum $\Spin$ 
of a black hole mentioned  
at the end of Sec.~\ref{sec:ID}, which is given by~\cite{Ashtekar2003},
\begin{equation}
\label{eq:angularmom}
\Spin = -\frac{1}{8\pi}\int_{\mathcal{S}_v} \! \phi^{\mu}\omega_{\mu} \, \sqrt{\bar{q}}d^2 x,
\end{equation}
for any choice of rotation vector field $\phi^\mu$ on $\mathcal{S}_v$. 
Because of this relation, this term has been interpreted 
as a flux of rotational energy~\cite{Ashtekar2003,Schnetter2006}.
However, it has been pointed out in~\cite{BoothFairhurst2007} that 
this is unlikely, as $\omega_\mu$ is related to 
$\Spin$ itself and not its flux. Indeed, this may be illustrated by 
considering a Kerr black hole that is distorted by an ingoing 
spherically symmetric dust shell (which carries no angular momentum). 
So even though there will be no flux of rotational energy, the last term 
in Eq.~\eqref{eq:DHfluxlaw} will still be nonzero whenever $\bar{C}\neq0$, 
which is necessarily true on a dynamical horizon. 
This last term also closely resembles the extremality parameter $e$ 
mentioned in Sec.~\ref{sec:DH}.

Another interpretation of the last term in Eq.~\eqref{eq:DHfluxlaw}
has been given by Hayward~\cite{Hayward2006} as a flux of
longitudinal gravitational radiation, by examining the components of
an effective gravitational radiation energy tensor in spin-coefficient form.
At future null infinity, the outgoing longitudinal gravitational
radiation is negligible relative to the outgoing transverse radiation,
but near the black hole this is generally not so.

To evaluate the dynamical horizon flux law, we first construct a 
tangent vector $X^\mu$ to $\mathcal{H}$ that connects $\mathcal{S}_v$ 
in $\Sigma_t$ to $\mathcal{S}_{v'>v}$ in $\Sigma_{t'}$ as 
\begin{equation}
\label{eq:initialDHtangent}
X^{\mu} = \pm\left(1,\frac{\partial x^i_v}{\partial t}\right),
\end{equation}
where $x^i_v$ are the coordinates of $\mathcal{S}_v$, and the plus 
sign is for $t'>t$ while the minus sign is for $t'<t$. 
The latter occurs along MTT2 and MTT4.  
The spatial components of the tangent vector $X^{\mu}$ 
diverge when two MTT sections meet. 
This may be avoided by a different choice 
of $X^{\mu}$, but here we employ the simple one described above. 
For this reason, we also consider the corresponding foliation 
parameter $v$ along each section of $\mathcal{H}$ separately.
Since
\begin{equation}
{\cal L}_X v = \pm \frac{\partial v}{\partial t},
\end{equation}
and we would like this to be unity, 
it follows that $v=\pm t + v_0$, where $v_0$ is some constant 
along each MTT section.
We choose $v=t$ along MTT1. Along the other MTT sections, 
we choose $v_0$ so that $v=0$ on the first 
$\mathcal{S}_v$ we find on those sections.

Next we make $X^\mu$ orthogonal to $\mathcal{S}_v$ to obtain $V^\mu$ 
(while leaving the time component unchanged, so Eq.~\eqref{eq:foliationparam} 
is still satisfied with the choice of $v$ described above). 
To achieve this, we use the unit tangent vectors to $\mathcal{S}_v$,
\begin{align}
\label{eq:MOTStangents}
p^{\mu} = N_p\left(0,\frac{\partial x^i_v}{\partial \theta}\right)%
\hspace{2 mm}\text{and}\hspace{2 mm}%
q^{\mu} = N_q\left(0,\frac{1}{\sin\theta}\frac{\partial x^i_v}{\partial \phi}\right).
\end{align}
Here, $x^i_v(\theta,\phi)=c^i_{\rm MOTS}+%
r_{\rm MOTS}(\theta,\phi)d^i(\theta,\phi)$ where 
$r_{\rm MOTS}(\theta,\phi)$ is given in Eq.~(\ref{eq:MOTS}) and 
$d^i$ is the coordinate unit vector pointing from the origin  
$c^i_{\rm MOTS}$ of the expansion along the $(\theta,\phi)$-directions.  
Also, $N_p$ and $N_q$ are normalization factors such that $p^2=q^2=1$. 
Orthogonalizing $q^{\mu}$ against $p^{\mu}$ gives the vector
\begin{equation}
\label{eq:Q}
Q^{\mu} = N_Q\left(q^{\mu}-p^{\nu}q_{\nu}p^{\mu}\right),
\end{equation}
where $N_Q$ is again a normalization factor such that $Q^2=1$. 
Then we obtain the desired tangent vector to $\mathcal{H}$ as 
\begin{align}
\label{eq:DHtangent}
V^{\mu} &= X^{\mu}-\left(p^{\nu}X_{\nu}\right)p^{\mu}%
                  -\left(Q^{\nu}X_{\nu}\right)Q^{\mu}.
\end{align}
This can be also be expressed in terms of our standard 
null normals of Eq.~\eqref{eq:nullnormals} as
\begin{equation} 
V^{\mu} = Bl^{\mu}-Ck^{\mu}, 
\end{equation}
with coefficients $B=-V^{\mu}k_{\mu}$ and $C=V^{\mu}l_{\mu}$.

\begin{figure}
\includegraphics[scale=0.5]{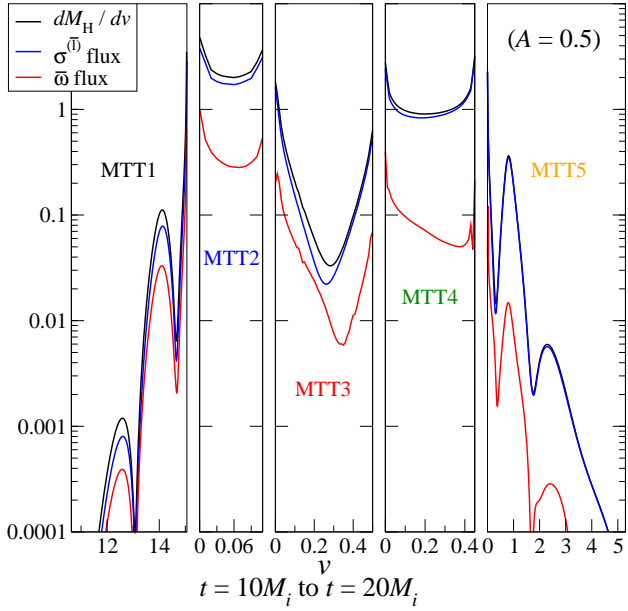}
\caption{
\label{fig:DHIrrMassFlux5}
Terms in the dynamical horizon flux law of Eq.~\eqref{eq:DHfluxlaw}
plotted against the foliation parameter $v$ along each section of
$\mathcal{S}_v$. Along MTT1, we choose $v=t$. Along the other MTT
sections, we choose $v=0$ on the first $\mathcal{S}_v$ we find.
}
\end{figure}

Now we determine the rescaled null normals $\bar{l}^{\mu}$ and 
$\bar{k}^{\mu}$ appearing in Eq.~\eqref{eq:DHtangentrescaled}. 
Since $V^{\mu}$ must be the same vector whether it 
is written in terms of $l^{\mu}$ and $k^{\mu}$, or 
$\bar{l}^{\mu}$ and $\bar{k}^{\mu}$, we have the relations 
\begin{align}
\label{eq:DHtangentcoefficients}
\bar{B}=B/f \hspace{2 mm}\text{and}\hspace{2 mm} \bar{C}=fC,
\end{align}
which together with Eq.~\eqref{eq:Cbar} gives
\begin{equation}\
\label{eq:scaling}
f = \frac{B}{\bar{B}} = \frac{\bar{C}}{C} 
  = \frac{2}{C}\frac{d M_{\text{H}}}{d v}.
\end{equation}
Evaluating the scale factor $f$ requires knowledge of  
$d M_{\text{H}}/dv$. It is straightforward to show 
that the area element $\sqrt{\bar{q}}$ of $\mathcal{S}_v$ changes  
along $\mathcal{H}$ as
\begin{equation}
{\cal L}_V \sqrt{\bar{q}} = -C\theta_{(k)}\sqrt{\bar{q}},
\end{equation}
so the change in the cross-sectional area $A_{\text{H}}$ along $\mathcal{H}$ is
\begin{equation}
\label{eq:areaflux}
\frac{d A_{\text{H}}}{d v} = -\int_{\mathcal{S}_v}%
  \! C\theta_{(k)} \, \sqrt{\bar{q}}d^2 x.
\end{equation}
From the definition $M_{\text{H}}=\sqrt{A_{\text{H}}/{16{\pi}}}$, 
it then follows that 
\begin{equation}
\frac{d M_{\text{H}}}{d v} = \frac{1}{\sqrt{64\pi A_{\text{H}}}}%
  \frac{d A_{\text{H}}}{d v}.
\end{equation}

The terms in the dynamical horizon flux law~\eqref{eq:DHfluxlaw} 
are calculated by noting that under the rescaling of the null 
normals $l^\mu$ and $k^\mu$,
\begin{align}
\label{eq:DHfluxtermsrescaled}
\sigma^{(\bar{l})}_{\mu\nu} = f\sigma^{(l)}_{\mu\nu}%
\hspace{2 mm}\text{and}\hspace{2 mm}%
\bar{\omega}_{\mu} = \omega_{\mu} + d_{\mu}\ln f.
\end{align}
The results are shown in Fig.~\ref{fig:DHIrrMassFlux5} from 
$t=10M_i$ to $t=20M_i$. The energy flux of matter is neglected 
since we have $T_{\mu\nu}=0$. The flux associated with 
$\bar{B}\sigma^{(\bar{l})}_{\mu\nu}\sigma^{(\bar{l})\mu\nu}$, 
labeled as ``$\sigma^{(\bar{l})}$ flux'',
is always the larger contribution 
to the growth of $M_{\text{H}}$, which is expected from the 
interpretation of this term as a flux of gravitational energy. 
This is most pronounced along MTT2 and MTT4, with decreasing $M_\text{H}$ 
during the evolution, and clearly indicates that 
their appearance is a consequence of the sufficiently high gravitational 
energy flux across them. We have seen in Sec.~\ref{sec:MultipleMTT} that 
for weak gravitational waves and with the same gauge condition for the 
evolution, no such MTTs appear.
The maximum number of MTTs that can exist at the same time may also 
be linked to the structure of the gravitational waves, as shown in 
the inset of Fig.~\ref{fig:EnergyMassRicciVsAmp}, although we have 
not explored this aspect further.

The fluxes increase rapidly near each bifurcation point.
This is because of our choice of 
normalization for $X^\mu$ in Eq.~(\ref{eq:initialDHtangent}), 
which propagates into $V^\mu$. 
To understand this, let us write as $x^{\mu}_c$ the spacetime coordinates of 
$\mathcal{S}_{c}$ that bifurcates, with foliation parameter 
$v=c$ say. Then on a nearby $\mathcal{S}_v$, we 
can approximate $\partial x^i_v / \partial t$ by
\begin{align}
\frac{\partial x^i_v}{\partial t} \approx%
  \frac{\partial}{\partial t} \left(x^i_c \pm \lambda\sqrt{|t-t_c}|\right)
  = \pm \frac{\lambda}{2}\frac{1}{\sqrt{|t-t_c|}},
\end{align}
where $\lambda$ is some function. As $t \rightarrow t_c$, this
quantity diverges as does the norm of $V^{\mu}$, and leads to the
higher values of the fluxes measured along $V^{\mu}$.  
This singular behavior could be absorbed into a redefined foliation
parameter $v'=v'(v)$.
Also, any visible discontinuities in the fluxes across different 
sections of $\mathcal{H}$ in Fig.~\ref{fig:DHIrrMassFlux5} are due to  
the difficulty in finding $\mathcal{S}_c$ exactly (as indicated 
by the data points in Fig.~\ref{fig:MOTSMass5}, even searching for 
MOTSs at every $\Delta t=0.01$ is insufficient for this purpose).

\subsection{Angular Momentum Flux Law}
\label{sec:AngMomFlux}

The angular momentum $\Spin$ defined in Eq.~\eqref{eq:angularmom}
depends on a choice of rotation vector $\phi^\mu$ on $\mathcal{S}_v$.
If $\mathcal{S}_v$ is axisymmetric, the natural choice of $\phi^\mu$
is the axial Killing vector. In general spacetimes no such
Killing vector exists, but one can nevertheless define a suitable
$\phi^\mu$~\cite{BoothFairhurst2005} by requiring it to have
closed orbits, and be divergence-free
\begin{equation}
\label{eq:DivFreePhi}
d_\mu \phi^\mu = 0.
\end{equation}
This notion has been further refined to calculate approximate Killing
vectors~\cite{Lovelace2008,Cook2007} in black hole simulations, and we will 
make use of this choice here. They were also used to compute 
$\Spin$ of the initial data sets in Sec.~\ref{sec:ID}.

Gourgoulhon has generalized the Damour-Navier-Stokes equation for null 
hypersurfaces to trapping horizons and used it to derive a flux law for the 
change in $\Spin$ along a hypersurface $\mathcal{H}$ 
foliated by 2-surfaces $\mathcal{S}_v$ (not necessarily MOTSs) with 
foliation parameter $v$~\cite{Gourgoulhon2005b},
\begin{align}
\label{eq:angularmomflux1}
\frac{d \Spin}{dv}%
  = &- \int_{\mathcal{S}_v} \! T_{\mu\nu}\phi^\mu \tau^\nu \, \sqrt{\bar{q}}d^2 x\\
  &- \frac{1}{16\pi} \int_{\mathcal{S}_v} \! \sigma^{(\tau)\mu\nu}{\cal L}_\phi \bar{q}_{\mu\nu} \, \sqrt{\bar{q}}d^2 x\notag\\
  &+ \int_{\mathcal{S}_v} \! \frac{1}{8\pi}\left[ \theta_{(k)}\phi^\mu d_\mu C - \omega_\mu {\cal L}_V \phi^\mu \right] \, \sqrt{\bar{q}}d^2 x\notag\\
\label{eq:angularmomflux2}
  = &- \int_{\mathcal{S}_v} \! T_{\mu\nu}\phi^\mu \tau^\nu \, \sqrt{\bar{q}}d^2 x\\
  &- \int_{\mathcal{S}_v}%
  \frac{1}{8\pi} \left[B\sigma^{(l)}_{\mu\nu}\sigma^{(\phi)\mu\nu}%
  + C\sigma^{(k)}_{\mu\nu}\sigma^{(\phi)\mu\nu} \right]%
  \, \sqrt{\bar{q}}d^2 x\notag\\
  &+ \int_{\mathcal{S}_v} \! \frac{1}{8\pi} \left[ \theta_{(k)}\phi^\mu d_\mu C - \omega_\mu {\cal L}_V \phi^\mu \right] \, \sqrt{\bar{q}}d^2 x\notag,
\end{align}
where the vectors $V^\mu=Bl^\mu-Ck^\mu$ and $\tau^\mu=Bl^\mu+Ck^\mu$ are 
tangent and normal to $\mathcal{H}$, respectively.  
The first integral in Eq.~\eqref{eq:angularmomflux2} is the
angular momentum flux due to matter. The second integral can be thought of 
as the flux due to gravitational radiation and vanishes if $\mathcal{S}_v$ 
is axisymmetric. In addition, it is usually required that $\phi^\mu$ be 
Lie transported along the dynamical horizon,
\begin{equation}
\label{eq:phicondition}
\mathcal{L}_V \phi^\mu = 0,
\end{equation}
so that the last integral in Eq.~\eqref{eq:angularmomflux2} 
vanishes when $\mathcal{S}_v$ is an MOTS~\cite{Gourgoulhon2005b}. 
This requirement ensures that in the 
absence of matter and gravitational radiation, the angular momentum flux will 
be zero along an MTT as expected, instead of there being some physically 
unmeaningful flux simply due to measuring $\Spin$ about different axes.  

\begin{figure}
\includegraphics[scale=0.5]{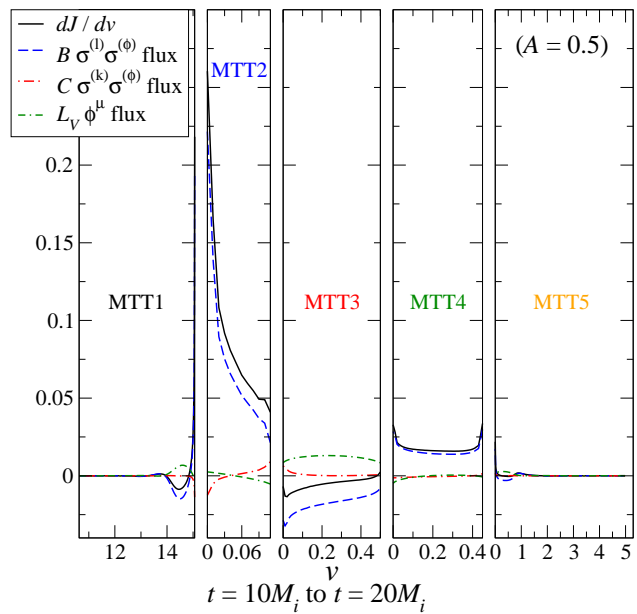}
\caption{
\label{fig:AngMomFluxAKV5}
Terms in the angular momentum flux law of Eq.~\eqref{eq:angularmomflux2}
plotted against the foliation parameter $v$ along each section of
$\mathcal{H}$. Along MTT1, we choose $v=t$. Along the other MTT
sections, we choose $v=0$ on the first $\mathcal{S}_v$ we find.
}
\end{figure}

Here we evaluate the angular momentum flux law for the dynamical horizon 
$\mathcal{H}$ found in Sec.~\ref{sec:DH} for $A=0.5$.
Because we calculate $\Spin$ with $\phi^\mu$ being an 
approximate Killing vector, Eq.~\eqref{eq:phicondition} is not satisfied 
in general, and so we must keep the last integral in 
Eq.~\eqref{eq:angularmomflux2}.
We use the same tangent vector $V^\mu$ and foliation parameter
$v$ along each section of $\mathcal{H}$ as in Sec.~\ref{sec:DHFlux},
and the null normals to $\mathcal{S}_v$ given in Eq.~\eqref{eq:nullnormals}. 
The values of the terms in Eq.~\eqref{eq:angularmomflux2} are shown in 
Fig.~\ref{fig:AngMomFluxAKV5} from $t=10M_i$ to $t=20M_i$.
The first integral is neglected since $T_{\mu\nu}=0$. 
The two terms in the second integral are labeled as 
``$B\sigma^{(l)}\sigma^{(\phi)}$ flux'' and 
``$C\sigma^{(k)}\sigma^{(\phi)}$ flux''. 
The last integral is labeled as ``$\mathcal{L}_V \phi^\mu$ flux''.
The angular momentum flux $d\Spin/dv$ is dominated by the flux 
associated with $B\sigma^{(l)}_{\mu\nu}\sigma^{(\phi)\mu\nu}$, due to the 
large $\sigma^{(l)}_{\mu\nu}$ produced by the gravitational waves. 
The magnitude of $d\Spin/dv$ vanishes initially, becomes largest along  
the end of MTT1 and the beginning of MTT2 when the gravitational waves reach 
the black hole, and settles back down to zero 
again along the successive MTT sections. Because $d\Spin/dv$ alternates 
sign along $\mathcal{H}$, the net change in $\Spin$ turns out to be 
small. The terms in the angular momentum flux law also diverge near each 
$\mathcal{S}_v$ that bifurcates into two MTTs, just like the terms in 
the dynamical horizon flux law in Fig.~\ref{fig:DHIrrMassFlux5}, and  
again is a consequence of our choice of $V^\mu$ as discussed at the end of 
Sec.~\ref{sec:DHFlux}.

\section{The Event Horizon}
\label{sec:EH}

\subsection{Basic Definitions and Concepts}
\label{sec:EH_Def}
The standard definition of the surface of a black hole is the 
$\textit{event horizon}$, the boundary of the set of all points 
that are not in the causal past of future null infinity~\cite{Wald}. 
It is a null hypersurface, generated by null geodesics that have no 
future endpoints. As defined, the event horizon is a 3-surface, but it is 
common to refer to the intersection of this surface with $\Sigma_t$ as the 
event horizon as well. In contrast to an MOTS, the event horizon can only be 
found after the entire future history of the spacetime is known. Because of its 
teleological nature, the event horizon can behave nonintuitively. For 
instance, before a gravitational collapse has occurred an event horizon 
already forms, even though there is no flux of energy or angular momentum 
across it yet. In this section we describe our method of finding the event 
horizon, and contrast its properties with those of the MTTs found in 
Sec.~\ref{sec:MTT}.

\subsection{Event Horizon Finder}
\label{sec:EH_Finder}
The event horizon is located in a spacetime by following geodesics
backward in time.  It is well known~\cite{Libson95a,Libson96} that
null outgoing geodesics in the vicinity of the event horizon, when
followed backwards in time, will converge onto the event horizon  
exponentially.  Therefore, given a well-chosen congruence of
geodesics, one can trace the event horizon of the spacetime with
exponentially (in time) improving accuracy.

Our event horizon finder~\cite{CohenPfeiffer2008} tracks a set of geodesics
backwards in time.  The initial guess for the event horizon is chosen
at some late time when the black hole is in a quasistationary
state.  At this time, the apparent horizon and event horizon coincide
closely, and the apparent horizon is used as the initial guess. 
The initial direction of the geodesics is chosen to be normal to the 
apparent horizon surface, and the geodesics are integrated backwards in 
time. The geodesic equation requires values for the metric and its
derivatives for each geodesic at each point in time.  These values are
obtained by interpolation from the values computed during the evolution.
With an appropriate form of the geodesic equation, 
we can follow a geodesic as a function of coordinate time $t$, rather 
than the affine parameter along the geodesic.

\begin{figure}
\includegraphics[scale=0.5]{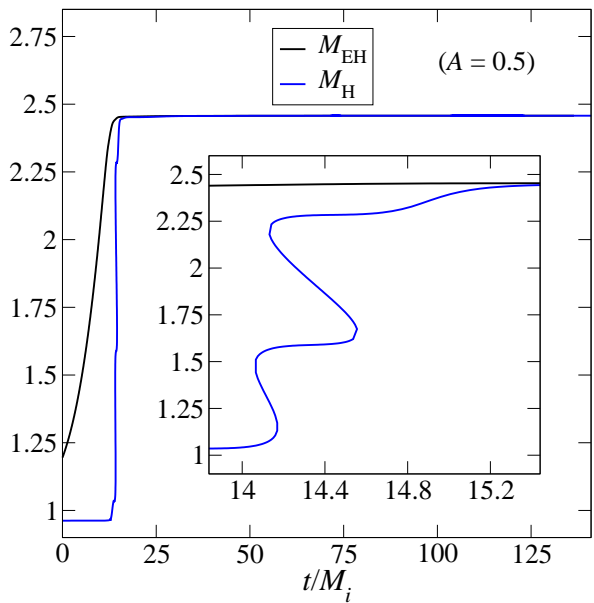}
\caption{
\label{fig:EHMOTS_Mass}
Irreducible masses of the event horizon $M_{\text{EH}}$ and the MOTSs
$M_{\text{H}}$ during the evolution with $A=0.5$.
At the very beginning of the evolution $M_{\text{EH}}$ is already increasing,
while $M_{\text{H}}$ is still fairly constant. As the inset shows,
$M_{\text{EH}}$ grows very slightly when $M_{\text{H}}$ changes the most.
}
\end{figure}

\subsection{Contrasting the Event Horizon with MTTs}
\label{sec:EHvsMTT}
We find the event horizon for the evolution in which the ingoing gravitational 
waves have the largest amplitude $A=0.5$.
The surface area $A_{\text{EH}}$ of the event horizon 
is computed by integrating the metric induced on its surface by the spatial 
metric $g_{ij}$. The irreducible mass of the event horizon is then given as  
$M_{\text{EH}}=\sqrt{A_{\text{EH}}/16\pi}$. This is shown in 
Fig.~\ref{fig:EHMOTS_Mass}, together with the irreducible mass $M_{\text{H}}$ 
along the MTTs. An obvious difference is that 
$M_{\text{EH}}$ always increases in time, and the event horizon does not 
bifurcate like the MTTs shortly after $t=14M_i$. 
The event horizon is also already growing at the very beginning of the 
evolution, before the gravitational waves have hit the black hole. By 
$t=14M_i$, the value of $M_{\text{EH}}$ has almost doubled while 
$M_{\text{H}}$ is still fairly close to its initial value. 
In fact, during the time when multiple MTTs are present and one would 
intuitively expect the black hole to be the most 
distorted, the event horizon shows very little growth. 

This peculiar behavior of the event horizon 
was also illustrated in~\cite{Booth2005} for the gravitational collapse of 
spherical dust shells, and explained with the null Raychaudhuri 
equation~\cite{PoissonToolkit},
\begin{equation}
\label{eq:NullRaychaudhuri}
\frac{d \theta_{(l)}}{d \lambda} = -\frac{1}{2}\theta_{(l)}^2%
  -\sigma^{(l)}_{\mu\nu}\sigma^{(l)\mu\nu}-8\pi T_{\mu\nu}l^{\mu}l^{\nu},
\end{equation}
where $\lambda$ is an affine parameter along the congruence of null geodesics 
that generate the event horizon, with tangent vector $l^{\mu}$.
The area element $\sqrt{h}$ of the event horizon is related to the 
expansion $\theta_{(l)}$ by $d\sqrt{h}/d\lambda=\theta_{(l)}\sqrt{h}$, 
and substituting this into Eq.~\eqref{eq:NullRaychaudhuri} gives
\begin{equation}
\label{eq:NullAreaEvolution}
\frac{d^2 \sqrt{h}}{d \lambda^2} = \left(\frac{1}{2}\theta_{(l)}^2%
  -\sigma^{(l)}_{\mu\nu}\sigma^{(l)\mu\nu}-8\pi T_{\mu\nu}l^{\mu}l^{\nu}%
  \right)\sqrt{h}.
\end{equation}
In dynamical situations we will generally have $\theta_{(l)}\neq0$ on the 
event horizon, and this accounts for its accelerated growth, which is 
evident even at early times in our evolution when the shear 
$\sigma^{(l)}_{\mu\nu}$ is negligible. When the pulse of gravitational 
waves hits the black hole, $\sigma^{(l)}_{\mu\nu}$ on the event horizon 
becomes large, and according to Eq.~\eqref{eq:NullAreaEvolution} this will 
decelerate its growth, even causing the growth to become very small in 
our case.

\begin{figure}
\includegraphics[scale=0.7]{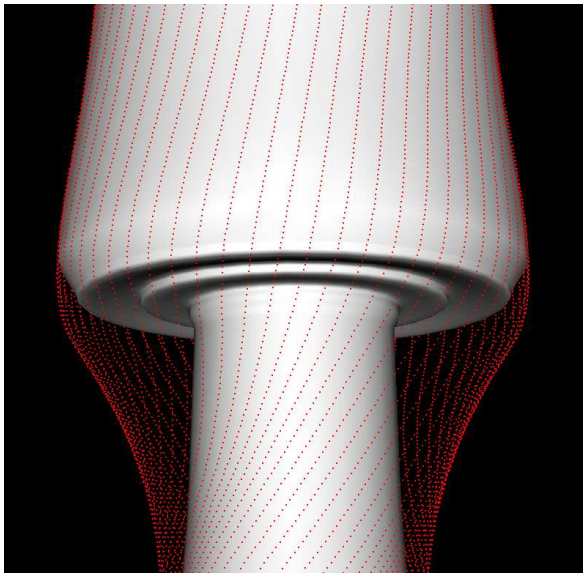}
\caption{
\label{fig:MTTs}
Spacetime diagram of the event horizon and dynamical horizons for
$A=0.5$. The dotted red lines are the null generators of the event horizon,
while the solid grey surface represents the dynamical horizons.
}
\end{figure}

At late times, the event and apparent horizons eventually coincide as 
both $\sigma^{(l)}_{\mu\nu}$ and $\theta_{(l)}$ go to zero 
on the event horizon while the apparent horizon becomes null.
Finally, Fig.~\ref{fig:MTTs} shows a spacetime diagram 
of the event horizon and the dynamical horizon $\mathcal{H}$, with the 
spatial dimension  along the $z-$direction suppressed. 
The null generators of the event horizon are shown as dotted red lines, 
and lie outside the solid grey surface of $\mathcal{H}$, 
except when they coincide at late times. 
In Fig.~\ref{fig:MTTs} the event horizon's cross section appears to be 
shrinking at late times.  The constancy of the area of the event horizon 
(cf. Fig.~\ref{fig:EHMOTS_Mass}) shows that this is merely a 
coordinate effect.

\section{Discussion}
\label{sec:Discussion}

In this paper, we investigate marginally trapped tubes and the event horizon 
for rotating black holes distorted by a pulse of ingoing gravitational waves.  
For small distortions (low amplitude $A$), 
the simulations do not exhibit any unexpected behavior: 
the area of the apparent 
horizon is initially approximately constant, it grows when the
gravitational radiation reaches the black hole, and then settles down
to a constant value after the highly dynamical regime is over.
However, for strong distortions, 
we find much more interesting behaviors of the MOTSs.  
A new pair of MOTSs appears outside the original MOTS.  These
new surfaces are initially close together and move rapidly away from
each other, indicating that at the critical time when they first
appear they are coincident (although this particular event cannot be
resolved in an evolution with finite time step).  The inner surface of
such a pair shrinks, eventually approaches the original MOTS, and then these 
two surfaces annihilate each other.
For amplitude $A=0.4$ this process happens once, for $A=0.5$ this happens
{\em twice}, and there is a short time interval during which 
{\em  five} MOTSs are present in the simulation. 

The MTTs traced out by the MOTSs are smooth, and appear to combine 
into one smooth hypersurface (although the
critical points where different marginally trapped tubes combine
with each other cannot be resolved).  When the black hole is 
distorted, we find that this hypersurface is 
everywhere spacelike and a dynamical horizon.  
We investigate how the black hole grows by evaluating the dynamical 
horizon flux law of Ashtekar and 
Krishnan~\cite{Ashtekar2003,BoothFairhurst2007}, and find that the 
gravitational energy flux is largest across the sections of the dynamical 
horizon that decrease in cross-sectional area with increasing time. 
We also evaluate the angular momentum flux law of 
Gourgoulhon~\cite{Gourgoulhon2005b} along the dynamical horizon, but 
instead of using a rotation vector $\phi^\mu$ that is Lie transported 
along the dynamical horizon, we use an approximate Killing 
vector~\cite{Lovelace2008}, since we prefer to calculate 
the angular momentum itself in this way. 
The angular momentum flux law is based on the generalized 
Damour-Navier-Stokes equation, which treats the black hole as a viscous 
fluid. Evaluating the generalized Damour-Navier-Stokes equation itself 
could aid in developing physical intuition about black holes in numerical 
spacetimes.

In illustrating the procedure for finding multiple MOTSs, caution 
must be taken to locate the apparent horizon with MOTS finders when the 
MOTS found at a previous time is used as an initial 
guess. If the MOTS finder is not run frequently enough, new MOTSs will be 
missed and an erroneous apparent horizon will be identified. 
This raises the issue of whether the true apparent horizon was indeed 
located in similar work involving highly distorted black holes in the 
past (e.g.~\cite{AnninosEtAl:1994}).
A better understanding of the slicing dependence
of the MOTSs in our simulations would also be helpful in choosing 
a more natural slicing condition that gives a single dynamical horizon 
that only grows in the cross-sectional area with time in highly dynamical 
situations.

When computing the event horizon, we find it to be smooth, and
enveloping the complicated structure of the MOTSs.  As can be seen in 
Figs.~\ref{fig:EHMOTS_Mass}
and~\ref{fig:MTTs}, the event horizon is very close to the apparent
horizon at late times, as one would expect.  The motion of the event
horizon is restricted by the fact that it is foliated by
null geodesics.  Therefore, in order to encompass the MOTSs, 
the event horizon begins to grow much earlier, and even at the start of our
simulation the event horizon is already considerably larger than the
apparent horizon.  At early times, $t\lesssim 10M_i$, 
the event horizon approaches the apparent horizon exponentially.  The rate of 
approach should be given by the surface gravity of the initial black hole, but
we have not verified this in detail, as our simulation does not reach
sufficiently far into the past. This could be checked by placing the initial 
pulse of gravitational radiation at a larger distance from the black hole.
The growth of the event horizon is described by the Hawking-Hartle 
formula~\cite{HawkingHartle1972}, which may also be evaluated to give 
a more complete comparison of MTTs and the event horizon. 

Our findings are analogous to the behavior of MOTSs and event horizons
in the Vaidya spacetime, as worked out in detail in the Appendix.  In
particular, for strong accretion, the Vaidya spacetime can also
exhibit multiple MOTSs at the same time, all of which foliate 
dynamical horizons.
Both in the Vaidya spacetime and our distorted Kerr spacetimes, the event
horizon begins to grow much earlier before multiple MOTSs appear.
By choosing a mass functions $m(v)$ that has two strong pulses of accretion, 
the Vaidya example in the Appendix would also produce five
concentric MOTSs similar to that seen in Fig.~\ref{fig:MOTSMass5}.


\acknowledgements{We thank Ivan Booth, Yanbei Chen, Stephen Fairhurst, 
  and Lee Lindblom for useful discussions. We are especially grateful 
  to Mark A. Scheel and Keith D. Matthews for discussions related to the 
  evaluation of the flux laws. 
  Calculations have been performed using the Spectral Einstein Code 
  ({\tt SpEC})~\cite{SpECwebsite}.
  This research was
  supported in part by grants from the Sherman Fairchild Foundation
  and the Brinson Foundation to Caltech and by NSF Grants
  No. PHY-0601459 and No. PHY-1005655 and NASA Grant No. NNX09AF97G at
  Caltech.  H.P. gratefully acknowledges support from the NSERC of
  Canada, from the Canada Research Chairs Program, and from the Canadian
  Institute for Advanced Research.  }

\appendix*
\section{Multiple Horizons in the Vaidya Spacetime}

The ingoing Vaidya spacetime is a spherically symmetric spacetime describing 
a black hole that accretes null dust~\cite{Vaidya1951}. It shares similar 
features to the distorted Kerr spacetimes presented in this paper, which 
we mention here briefly. The ingoing Vaidya metric in ingoing 
Eddington-Finkelstein coordinates $(v,r,\theta,\phi)$ is
\begin{equation}
\label{eq:Vaidya}
ds^2 = -\left(1-\frac{2m(v)}{r}\right)dv^2 + 2dvdr+r^2d\Omega^2,
\end{equation}
where $v=t+r$ is advanced time (not to be confused with the foliation 
parameter $v$ of dynamical horizon in the main text). 
From the Einstein equations, the stress-energy tensor is
\begin{equation}
\label{eq:VaidyaStressTensor}
T_{\mu\nu} = \frac{dm/dv}{4\pi r^2} (\partial_\mu v)(\partial_\nu v).
\end{equation}
With the choice of radial outgoing and 
ingoing null vectors 
\begin{equation}
\label{eq:VaidyaNulls}
l^\mu = \left[1,\frac{1}{2}\left(1-\frac{2m(v)}{r}\right),0,0\right]%
\hspace{2 mm}\text{and}\hspace{2 mm}%
k^\mu = (0,-1,0,0)
\end{equation}
normalized so that $l^\mu k_\mu=-1$, the expansions of the null normals are 
\begin{equation}
\label{eq:VaidyaExpansions}
\theta_{(l)} = \frac{1}{r}\left(1-\frac{2m(v)}{r}\right)%
\hspace{2 mm}\text{and}\hspace{2 mm}%
\theta_{(k)} = -\frac{2}{r}.
\end{equation}

\begin{figure}
\includegraphics[scale=0.5]{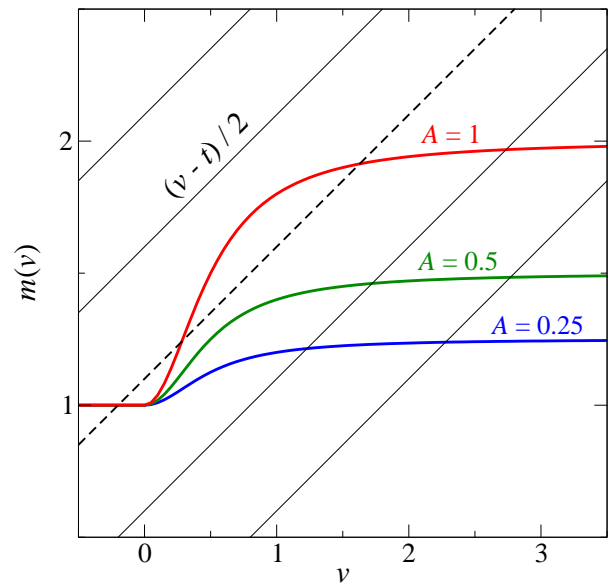}
\caption{
\label{fig:VaidyaIllustration}
Mass functions $m(v)$ of the Vaidya spacetime for three amplitudes
$A=0.25,0.5$, and 1, along with the straight lines $(v-t)/2$.  MOTSs
exist at the intersections of these functions. For $A=0.5$ and 1,
there are up to three intersections, as illustrated by the dashed
black line which intersects the $A=1$ mass curve three times.  }
\end{figure}

From this, we see that MOTSs are located at $r=2m(v)$,
or
\begin{equation}
\label{eq:VaidyaMOTSs}
m(v)=\frac{1}{2}(v-t).
\end{equation}
The number of solutions to Eq.~(\ref{eq:VaidyaMOTSs}), i.e. the number
of MOTSs, can be conveniently discussed with the diagram shown in
Fig.~\ref{fig:VaidyaIllustration}.  The thick solid lines represent
three different mass functions $m(v)$ plotted vs $v$.  The
right-hand side of Eq.~(\ref{eq:VaidyaMOTSs}) is a family of straight
lines (one for each $t$) represented by the thin diagonal lines in
Fig.~\ref{fig:VaidyaIllustration}.  For a given $t$, the number of
intersections between the $(v-t)/2$ and the $m(v)$ curve gives the
number of MOTSs at that particular $t$.  The straight line
$\frac{1}{2}(v-t)$ has slope 1/2, so if $dm/dv<1/2$ for all $v$, then
there will be exactly one intersection\footnote{Assuming $m(v)$ is
  non-decreasing, and has finite bounds for $v\to\pm \infty$.} for
every $t$. If
\begin{equation}\label{eq:MultiMOTS}
\frac{dm}{dv}>\frac{1}{2}\qquad\mbox{for some $v$,}
\end{equation}
then the $m(v)$ curve will have regions that are steeper than the
straight line. By adjusting the vertical intercept of the straight
line, equivalent to choosing a suitable $t$, the straight line will
pass through a point with $dm/dv>1/2$. At this point, $m(v)$ passes
from below to above the straight line, so there must be an additional
intersection at both smaller and larger $v$, for a total of three
MOTSs.  Thus, sufficiently rapid mass accretion (large $dm/dv$)
results in multiple MOTSs.

The signature of a spherically symmetric MTT depends on the sign 
of~\cite{Booth2006}
\begin{equation}
\label{eq:C}
C = \frac{T_{\mu\nu} l^\mu l^\nu}{1/(2A_{\rm H}) - T_{\mu\nu}l^\mu k^\nu}, 
\end{equation} 
where $A_{\rm H}$ is the cross-sectional area of the MTT. The MTT is spacelike if 
$C>0$, null if $C=0$, and timelike if $C<0$. From  
Eq.~\eqref{eq:VaidyaStressTensor} and Eq.~\eqref{eq:VaidyaNulls},
\begin{align}
T_{\mu\nu} l^\mu l^\nu = \frac{dm/dv}{4\pi r^2}%
\hspace{2 mm}\text{and}\hspace{2 mm}%
T_{\mu\nu}l^\mu k^\nu = 0,
\end{align}
so we see that $C>0$ for the Vaidya spacetime as long as $dm/dv>0$. 
Furthermore, since $\theta_{(k)}<0$, these MTTs will also be  
dynamical horizons. 

The event horizon is generated by radial outgoing null geodesics 
satisfying
\begin{equation}\label{eq:Vaidya-EH}
\frac{dr}{dv} = \frac{1}{2}\left(1-\frac{2m(v)}{r}\right).
\end{equation}
Integrating this differential equation requires knowledge of the 
event horizon location at some point. This is usually supplied by 
the final state of the black hole, when accretion has ended.

\begin{figure}
\includegraphics[scale=0.5]{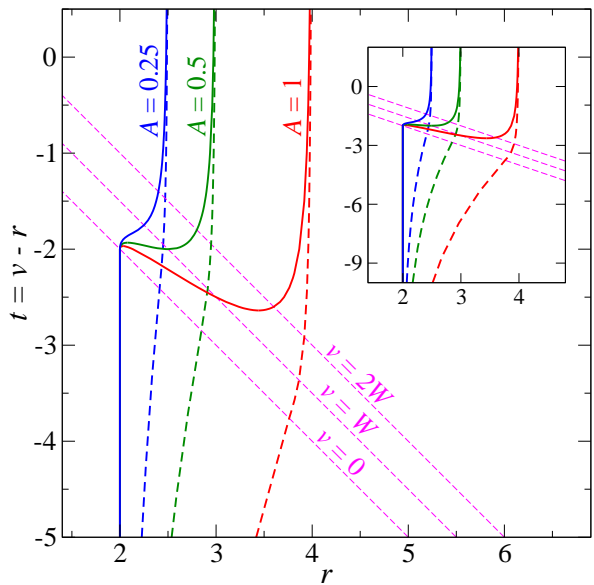}
\caption{
\label{fig:VaidyaHorizons}
Locations of MOTSs (solid lines) and event horizons (dashed lines) 
in the Vaidya spacetime. For
$A=0.25$ there is one MOTS at all times. For $A=0.5$ and 1, up to
three MOTSs exist at a time $t$.  The event horizons approach the MTTs
at very early and late times, and start growing much earlier than the
MTTs.  The inset shows a larger interval in $t$.}
\end{figure}

To close, we illustrate these considerations with a concrete example. 
We choose the mass function
\begin{equation}
\label{VaidyaMass}
m(v)=\left\{\begin{aligned} &m_0,&\qquad v\leq0\\ 
      &m_0+\frac{Am_0 v^2}{v^2 + W^2},&\qquad v>0\end{aligned}\right.
\end{equation}
similar to that presented in~\cite{Schnetter2006b} ($Am_0$ is the mass
accreted by the black hole, and $W$ determines the time scale of
accretion).  We set $m_0=1$, $W=0.5$, and consider three different
amplitudes $A=0.25,0.5$, and 1.  Figure~\ref{fig:VaidyaIllustration} shows
the respective mass functions, and we see that $A=0.25$ never leads to
multiple MOTSs, while $A=1$ clearly exhibits three MOTSs for certain
$t$.  It is easy to show that Eq.~(\ref{eq:MultiMOTS}) implies $Am_0 >
4W/(3\sqrt{3})$. The locations of the MOTSs in $(r,t)$ coordinates are shown in
Fig.~\ref{fig:VaidyaHorizons}. For $A=0.25$, there is only one MOTS at
all times. For $A=0.5$, there are up to three MOTSs at a single time.
A new MOTS appears at $r=2.5$ immediately after $t=-2$, and bifurcates
into two MTTs. One of these MTTs shrinks and annihilates with the
innermost MTT at $t=-1.93256$, while only the outermost MTT remains at
late times and grows towards $r=3$. For $A=1$, there are again up to
three MOTSs at a single time, but a new MOTS appears earlier at
$t=-2.63822$.  After $t=-1.96824$, only one MOTS remains and grows
towards $r=4$.  Also shown in Fig.~\ref{fig:VaidyaHorizons} are lines
of constant $v$ indicating when accretion begins ($v=0$), and when $m(v)$
has increased by $50\%$ and $80\%$, respectively ($v=W$ and $v=2W$).

The event horizons for the three cases are computed by integrating
Eq.~(\ref{eq:Vaidya-EH}) {\em backward} in time, starting with $r_{\rm
  EH}(v\to\infty)=2(1+A)m_0$. The resulting surfaces are shown as the
dashed curves in Fig.~\ref{fig:VaidyaHorizons}.  The event horizon is
located at $r=2$ in the far past, starts growing long before $m(v)$
increases, and asymptotically approaches the MTT of the final black
hole for all amplitudes $A$.

\bibliography{References/References}
\end{document}